\documentclass[preprint,authoryear,12pt]{elsarticle}

\usepackage{xcolor}
\usepackage{amsmath, amssymb, amsthm, eucal, multicol}
\usepackage{gensymb}
\usepackage{booktabs}
\usepackage{subfig}

\newenvironment{sistema}%
{\left\lbrace\begin{array}{@{}l@{}}}%
{\end{array}\right.}

\newcommand{\de}{\mathrm{d}}

\newcommand{\Ue}{U_\epsilon}
\newcommand{\Ta}{T_\alpha}
\newcommand{\Tb}{T_\beta}
\newcommand{\Ca}{C_\alpha}
\newcommand{\Cb}{C_\beta}

\journal{Journal of Volcanology and Geothermal Research}

\begin{document}

\begin{frontmatter}



\title{Ash plume properties retrieved from infrared images: a forward and inverse modeling approach}


\author[sns,ingv]{Matteo Cerminara\corref{cor1}\fnref{fn1}}
\ead{matteo.cerminara@gmail.com}
\cortext[cor1]{Corresponding author. Tel.: +393381272966.}
\fntext[fn1]{Present address: Istituto Nazionale di Geofisica e Vulcanologia, Sezione di Pisa, via
della Faggiola 32, 56126, Pisa, Italy.}
\author[ingv]{Tomaso Esposti Ongaro}
\author[unifi]{S\'ebastien Valade}
\author[unibp]{Andrew J.L. Harris}

\address[sns]{Scuola Normale Superiore di Pisa, Italy}
\address[ingv]{Istituto Nazionale di Geofisica e Vulcanologia, Sezione di Pisa, Italy}
\address[unifi]{University of Florence, Department of Earth Sciences, Italy}
\address[unibp]{Universit\'e Blaise Pascal, Laboratoire Magmas et Volcans, Clermont-Ferrand, France}

\begin{abstract}
We present a coupled fluid-dynamic and electromagnetic model for volcanic ash plumes. 
In a forward approach, the model is able to simulate the plume dynamics from prescribed input flow conditions and generate the corresponding synthetic thermal infrared (TIR) image, allowing a comparison with field-based observations. An inversion procedure is then developed to retrieve ash plume properties from TIR images. 

The adopted fluid-dynamic model is based on a one-dimensional, stationary description of a self-similar (top-hat) turbulent plume, for which an asymptotic analytical solution is obtained.
The electromagnetic emission/absorption model is based on the Schwarzschild's equation and on Mie's theory for disperse particles, assuming that particles are coarser than the radiation wavelength and neglecting scattering. 
In the inversion procedure, model parameters space is sampled to find the optimal set of input conditions which minimizes the difference between the experimental and the synthetic image.
Two complementary methods are discussed: the first is based on a fully two-dimensional fit of the TIR image, while the second only inverts axial data.
Due to the top-hat assumption (which overestimates density and temperature at the plume margins), the one-dimensional fit results to be more accurate. However, it cannot be used to estimate the average plume opening angle. Therefore, the entrainment coefficient can only be derived from the two-dimensional fit.

Application of the inversion procedure to an ash plume at Santiaguito volcano (Guatemala) has
allowed us to retrieve the main plume input parameters, namely the initial radius $b_0$, velocity
$U_0$, temperature $T_0$, gas mass ratio $n_0$, entrainment coefficient $k$ and their related
uncertainty. Moreover, coupling with the electromagnetic model, we have been able to
obtain a reliable estimate of the equivalent Sauter diameter $d_s$ of the total particle size
distribution.

The presented method is general and, in principle, can be applied to the spatial distribution of
particle concentration and temperature obtained by any fluid-dynamic model, either integral or
multidimensional, stationary or time-dependent, single or multiphase.
The method discussed here is fast and robust, thus indicating potential for applications to
real-time estimation of ash mass flux and particle size distribution, which is crucial for
model-based forecasts of the volcanic ash dispersal process.
\end{abstract}

\end{frontmatter}



\section{Introduction}
\label{intro}
Volcanic plumes are produced during explosive eruptions by the injection of a high-temperature gas-particle mixture in the atmosphere.
The dynamics of ascent of a volcanic plume injected into the atmosphere is controlled by several factors (including vent overpressure, crater shape, wind) but, following \citet{morton1956} and \citet{Wilson1976} it has been recognized that mass flow rate (or eruption intensity) and mixture temperature mainly control the final plume height in stratified environments. 
In convective regimes and for the most intense eruptions, volcanic plumes are able to reach stratospheric layers, where ash can persist for years and affect climate, mesoscale circulation, air quality and endanger aviation transports. Due to their associated fallout, even weak volcanic plumes can have large impacts on populations living close the the volcano, especially in highly urbanized regions around active
volcanoes. 

Despite the advancement of physical models describing eruption conditions and the subsequent atmospheric dispersal of the gas-particle mixture during an explosive eruption, one of the main obstacles to the full understanding of volcanic plume dynamics is the difficulty in obtaining measurements of the ascent dynamics and plume properties. Indeed, not only are measurements difficult and dangerous, but repeatability is an issue: eruptions differ from each other (even at the same volcano) and are too infrequent to allow construction of a statistically robust data set \citep{deligne2010recurrence}. 
While this is certainly true for large eruptions \citep[exceeding 0.1 km$^3$ of total erupted mass, or a Volcanic Explosive Index -- VEI $> 3$ --][]{newhall82vei}, persistent but low intensity explosive activity is common at several volcanoes such as Stromboli \citep{giberti1992steady,harris2007temperature}, Santiaguito \citep{bluth2004,johnson2004explosion,sahetapy2009} and Soufrière Hills \citep{druitt2002eruption,Clarke2002}.
Such systems offer natural laboratories where methodologies and models for application to rarer, but more energetic events, can be prepared and tested. However, even in such cases, direct measurement in syneruptive conditions is extremely difficult, so that volcanologists must rely on indirect (remote sensing) measurement techniques.

Our current understanding of volcanic plume dynamics is largely based on visual observations and on one-dimensional plume models. 
However, there is a general consensus that the fundamental mechanism driving the ascent of volcanic plumes is the conversion of its thermal energy into kinetic energy through the quasi-adiabatic expansion of hot volcanic gases and atmospheric air entrained by turbulence  \citep{sparks1997volcanic}. One-dimensional models translating this concept into mathematical language \citep{Wilson1976,woods1988} have played a key role in advancing our understanding of the physics of volcanic plumes. 
One of the reasons of their success is also that simple models rely on simple measurements for validation, allowing solution with a limited number of parameters.
In the case of eruption plume models, one observable is sufficient, namely plume height. 
This can be measured using photogrammetry, infrared imaging, satellite remote sensing, ceilometers, radio and radio-acoustic sounding  \citep{tupper2003observations}. 
Only one adjustable parameter is then needed to fit plume observations, namely a self-similarity
coefficient, or the entrainment coefficient. This linearly correlates the rate of entrainment to the
average vertical plume velocity \citep{ishimine2006}.

However, the plume interior is generally invisible to the observer, and there is no way to measure
mixture density from simple visual observation. As a result other electromagnetic imaging techniques
(here defined as the process by which it is possible to observe the internal part of an object which
cannot be seen from the exterior) are needed to obtain data regarding the plume interior
\citep{scollo2012monitoring}.

Forward looking infrared (FLIR) cameras have become affordable in the last years and their use in volcanic plume monitoring has become popular \citep{spampinato2011volcano,ramsey2012volcanology,Harris2013}. 
To date they have been used to classify and measure bulk plume properties, such as plume front ascent rates, spreading rates and air entrainment rates for both gas, ash and ballistic rich emissions \citep{harris2007temperature,patrick2007strombolian,sahetapy2009}, analysis of particle launch velocities, size distributions and gas densities \citep{harris2012detailed,delledonne2012} and particle tracking velocimetry \citep{bombrun2013particle}. 
Recent deployments have involved use of two thermal cameras: one close up to capture the at-vent dynamics as the mixture exits from the conduit and one standing off to obtain full ascent dynamics as the plume ascends to its full extent. 
This has been coupled with stereo-visible-camera, Doppler radar measurements of the same plume and infrasonic measurement, to detect large-scale puffing and eddies. Recently, \citet{valade2014} have developed a procedure to extract from thermal infrared (TIR) images an estimate of the entrainment coefficient and other plume properties including plume bulk density, mass, mass flux and ascent velocity.

However, recovery of the plume ash mass content and grain size distribution in near-real time remains a major challenge. 
Such data are crucial for hazard mitigation issues, and especially for the Volcanic Ash Advisory Centers (VAACs) which issue advisories to the aviation community during explosive eruptions. 
Indeed, VAACs use ash dispersion models (VATD, Volcanic Ash Transport and Dispersion models) to forecast the downstream location, concentration, and fallout of volcanic particles \citep{stohl2010}. 
However, to be accurate, such models require quantification of the plume ash concentration and particle size distribution \citep{mastin2009,bonadonna2012}. 

In this work we show that recovering this information is possible in a rapid and robust fashion by comparing thermal infrared images that record the emission of a volcanic plume, with synthetic thermal infrared images reconstructed from analytical models.

Our approach inverts time-averaged thermal image data to reconstruct the temperature, ash concentration, velocity profiles and the grain size distribution within the plume. 
To do this we construct a synthetic thermal image of the volcanic plume starting from the spatial distribution of gas and particles obtained from a fluid dynamic model. 
The method is based on the definition of the infrared (IR) irradiance for the gas-pyroclast
mixture. This is derived from the classical theory of radiative heat transfer \citep{modest2003}
with the approximation of negligible scattering (Schwarzschild's equation).
The model needs to be calibrated to account for the background atmospheric IR radiation and the material emissivity \citep{Harris2013}.
The absorption and transmission functions needed to compute the irradiance are derived from the Mie's theory \citep{mie1908} and can be related, by means of semi-empirical models, to the local particle concentration, grain size distribution and to the optical thickness of the plume. 
By applying such an IR emission model to the gas-particle distribution obtained from a fluid dynamic model it is possible to compute a synthetic thermal image as a function of the input conditions. 
We adopt a one-dimensional, time-averaged plume model derived from the \citet{woods1988} model to simulate the plume profile. 
The advantage of 1D modeling is that inversion can be performed in a fast and straightforward way by means of minimization of the difference between a synthetic and a measured IR image. However, the method is applicable to any kind of plume model.

\begin{figure}
\centering
\includegraphics[width= \columnwidth]{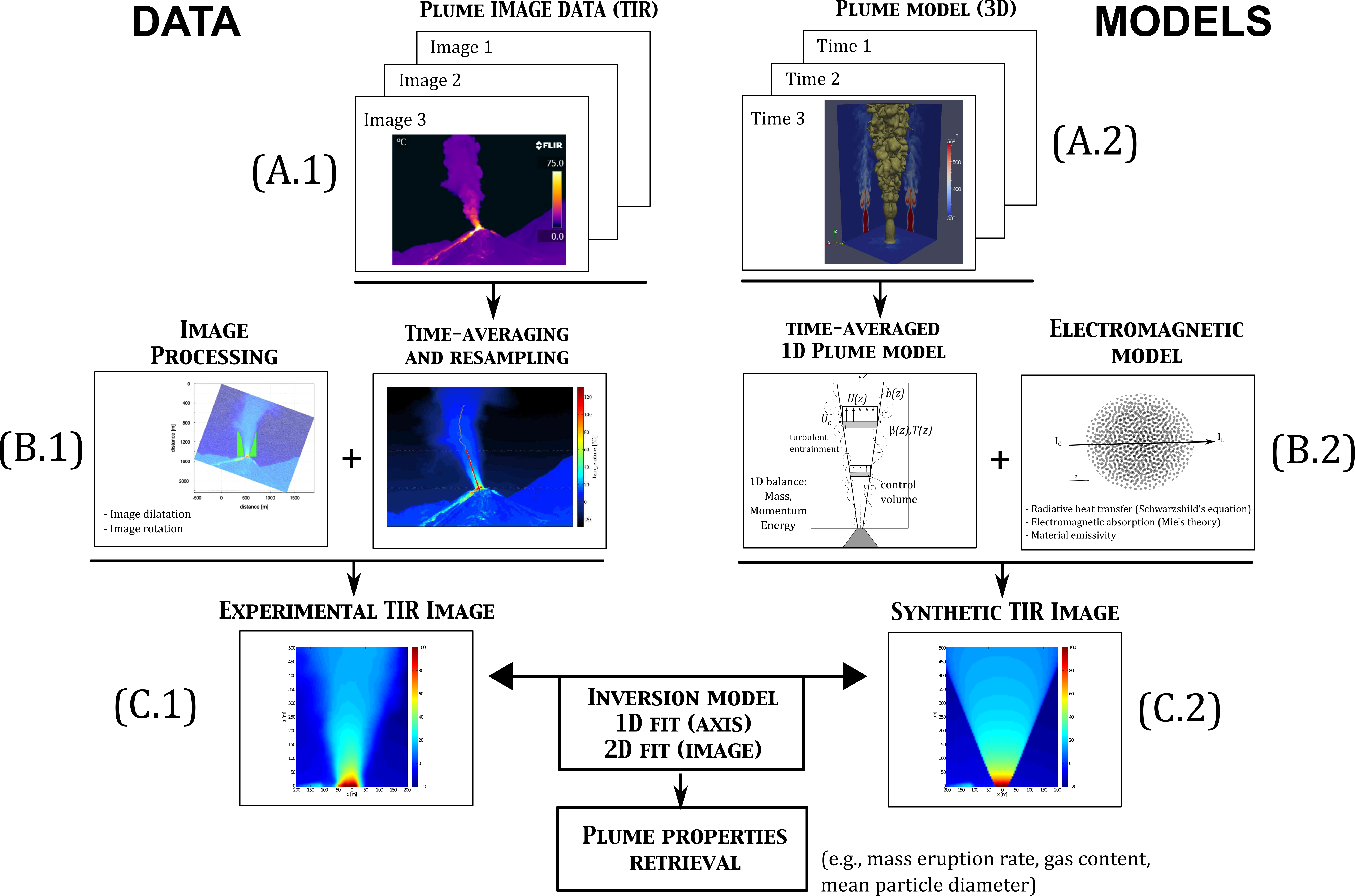}
\caption{Schematic overview of the aims and methodology of the work presented in this paper. (A.1)
Volcanic plume emission is recorded using a thermal infrared (TIR) camera. Modeling of such
phenomenon requires 3D numerical models (A.2), able to reproduce the complex fluid-dynamic behavior
at various length and time scales. However because such models require high computational power and
time, analytical 1D mean plume models (B.2) may be used to predict the mean behaviour of volcanic
plumes. Such models are time-averaged stationary models, which describe the mean spatial
distribution of flow parameters (e.g., particle concentration, temperature, velocity) given a set of
input conditions. By coupling an electromagnetic model to the plume model, we simulate the TIR
emission of the gas-particle mixture, and compute a synthetic thermal infrared image (C.2). The mean
plume behaviour may also be recovered from the recorded image by constructing a "mean image" (B.1),
which is a time-averaged image obtained from 
averaging a sequence of images in a TIR video sequence. In doing this, the time-dependent dynamic
fluctuations of the plume are filtered, leaving an image that reflects the mean plume behaviour.
Image processing is then applied to obtain an image with a vent-centered metric coordinate system,
comparable to that created by the forward model. Recursive minimization of the discrepancy between
the observed and modeled TIR images is then performed by application of an inversion model (2D when
the entire images are compared, or 1D when only the plume central axis is compared), which searches
for the best model input-parameters (e.g., ash mass, particle size distribution, etc.) that
reproduce the observed data.}
\label{fig:methodology}
\end{figure}

Thus, In Section~\ref{em} we present the IR electromagnetic model (equations and approximations)
that we use to produce plume synthetic images. In Section~\ref{model} we describe the
one-dimensional integral fluid-dynamic model of the plume. In Section~\ref{fluid-em} we apply the
coupled fluid-dynamic-electromagnetic model (forward model) to construct a synthetic thermal image
of a volcanic plume. In Section~\ref{inversion} we use this model to invert experimental TIR data
acquired during an explosive event at Santiaguito volcano (Guatemala) to estimate the flow
conditions at the vent.
Figure \ref{fig:methodology} illustrates the methodology and models developed in the paper.

\section{Electromagnetic model}
\label{em}
Due to the high-temperature of erupted gas and pyroclasts, volcanic plumes emit electromagnetic
radiation in the thermal infrared (TIR) wavelengths (8--14 $\mu$m).
Every single particle radiates as a function of its temperature (through the Planck's function) and
material properties (each material being characterized by its {\em emissivity} -- see
Sect.\ref{sect:emissivity}).
On the other hand, part of the emitted radiation is absorbed by neighbouring gas and particles, so that the net transmitted radiation results from the balance between emission and absorption and is a function of the electromagnetic wavelength $\lambda$. 
This balance is expressed by Schwarzschild's equation.
\subsection{Schwarzschild's equation}
\begin{figure}
\centering
\includegraphics[width=0.5 \columnwidth]{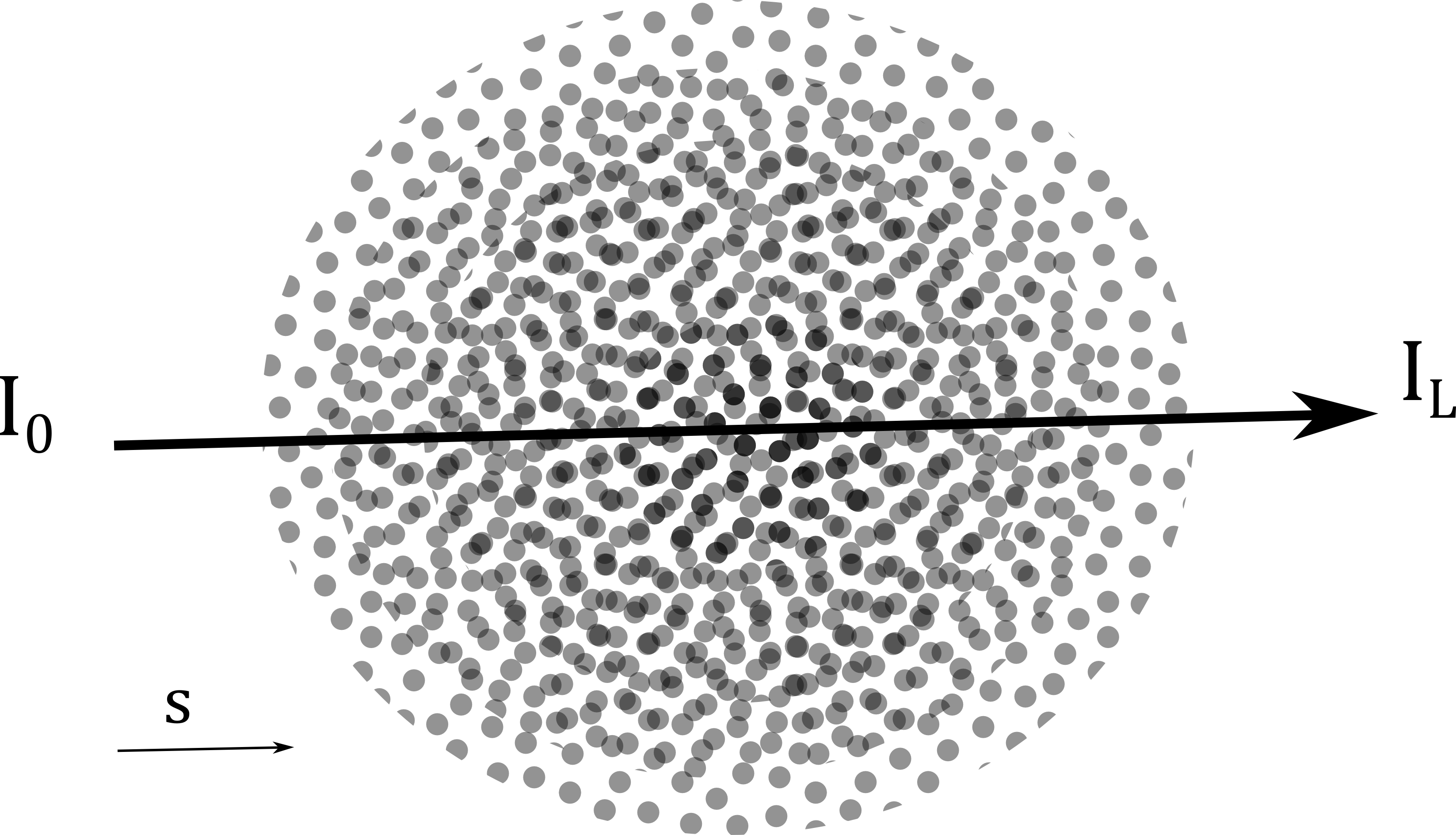}
\caption{Schematic configuration of the propagation of a TIR ray inside a heterogeneous medium. The
absorption coefficient is proportional to particle concentration. TIR intensity changes along
the path due to particle emission/absorption. We indicate with $s$ the curvilinear coordinate along
the path, with $I_0$ being the background intensity and $I_L$ being the measured intensity at distance $L$
from the background position.}
\label{fig:optical_path}
\end{figure}

Along an optical path, defined by a curvilinear coordinate $s$ (see Fig.~\ref{fig:optical_path}),
the infinitesimal variation of TIR intensity due to emission at temperature $T$ is proportional to
the Planck function $B=B_{\lambda}(T)$ multiplied by the infinitesimal length $ds$: $dI_{emit}=\beta_e B
ds$.
On the other hand, the infinitesimal variation due to absorption is proportional to the radiation
intensity itself, so that
$dI_{abs}=\beta_a I ds$.
Following Kirchoff's law, the emission and absorption coefficients are equal $\beta_e=\beta_a=K$, so that, along a ray, this balance is expressed by:
\begin{equation}
\frac{dI}{ds}=K(B-I)\,.
\label{eq:differentialSch}
\end{equation}
By solving Eq.\eqref{eq:differentialSch} along the optical path (as represented in Figure
\ref{fig:optical_path}), for a heterogeneous medium, we find that
\begin{equation}
I_{\lambda}(L) = I_0 e^{-\tau_L} + \int_0^L K_{\lambda}(s)B_{\lambda}(s) e^{-(\tau_L-\tau(s))}ds
\label{eq:Ilambda}
\end{equation}
where $I_0 = I(0)$ is the background atmospheric radiation at the given wavelength and the integral
is computed along a straight ray from the source $s=0$ to $s=L$, this being the detector position.
The {\em optical thickness} is defined as:
\begin{equation}
\label{eq:taus}
\tau(s)=\int_0^s K(s) ds
\end{equation}
and $\tau_L=\tau(L)$.
Because the medium is heterogeneous (i.e., it can have a non-homogeneous particle concentration and temperature) the absorption coefficient $K$ depends on the position $s$ along the optical path. 
In the next section, we show how $K(s)$ can be derived for a cloud of particles.
\subsection{Absorption coefficient of the particulate phase}
The absorption coefficient for a cloud of disperse spherical particles can be derived from Mie's
theory \citep{mie1908,hanel1977approximation}. Up to a limiting size $r_{\lambda}$, which depends on
the measurement wavelength \citep[see Fig. 1 of ][]{hanel1977approximation}, $K$ can be described by
a power law.
Above the upper particle size limit, absorption no longer depends on the particle size or material, but simply corresponds to the total cross section of the dispersed particles:
\[
K=\frac{1}{V}\sum_{j=1}^N \pi r_j^2
\]
where $r_j$ is the radius of the single sphere and $V$ is the total volume occupied by the mixture.
For the case of volcanic particles, this condition is almost always satisfied when particles are
much larger than the measurement wavelength (about 10 $\mu$m).

By expressing the volume $V$ in terms of the density and particle concentration $\varepsilon$, the absorption coefficient can be set propotional to the Sauter diameter $d_s$ of the particle distribution, i.e.
\begin{equation}\label{eq:k}
K=\frac{3}{4}\varepsilon
\frac{\sum_jr_j^2}{\sum_jr_j^3}=\frac{3}{2}\frac{\varepsilon}{\overline{d^3}/\overline{d^2}}=\frac{3
} { 2 } \frac{\varepsilon}{d_s}
\end{equation}
or, in terms of the particle microscopic density $\hat{\rho}_s$ and the particle bulk density ($\rho_s =
\varepsilon \hat{\rho}_s$)

\begin{equation}\label{eq:ks}
K_s = \frac{3}{2d_s \hat{\rho}_s}\, \rho_s = A_s \rho_s
\end{equation}

Here we have introduced the specific absorption coefficient of the particles $A_s$ which represents the cross section of the particulate phase per unit of mass (having the dimensions of [m$^2$/Kg]).
The Sauter diameter $d_s$ represents the mean particle diameter that gives the same volume/surface area ratio as the original particle size distribution.
For example, we can estimate the mean Sauter diameter in the case of a log-normal grain size
distribution. This is a common hypothesis in volcanology when we assume that the grain size
distribution is Gaussian in $\phi=-\log_2(d)$ units, with mean $\mu$ and standard deviation
$\sigma_{\rm gsd}$.
In such case, we can write the normalized particle distribution (in millimeters) as:
\[
f(x;\mu,\sigma_{\rm gsd}) = \frac{1}{x \ln(2) \sigma_{\rm gsd} \sqrt{2 \pi}} e^{-\frac{(\log_2(x)+\mu)^2}{2\sigma_{gsd}^2}}
\]
and the Sauter mean diameter can be computed analitically as
\begin{equation}
\label{eq:d_s}
d_s = 2^{-\mu + \frac{5}{2}\log(2)\sigma_{\rm gsd}^2}\,.
\end{equation}
This means that the mean particle diameter ``seen'' by the TIR sensor is always
larger than the mean diameter $\bar{d} = 2^{-\mu + \frac{1}{2}\log(2)\sigma_{\rm gsd}^2}$.

In the context of non-homogeneous mixtures, the particle volumetric fraction $\varepsilon$ in
Eq.\eqref{eq:k} changes with the position $s$ along the ray trajectory, i.e.,
$K=K(s)=\frac{3}{2}\frac{\varepsilon(s)}{d_s}$. If the grain size distribution changes locally, the
dependency of $d_s(s)$ should also be taken into account.

In Section \ref{fluid-em} we will detail how the optical thickness $\tau(s)$ and
Eq.\eqref{eq:Ilambda} are computed in volcanic plume applications.

\subsection{Emissivity of the particulate phase}
\label{sect:emissivity}

To estimate the emitted electromagnetic intensity, we assume that pyroclastic particles behave as {\em gray bodies}. 
The ratio between the emitted radiation and the black-body intensity at the same temperature (given
by the Planck function) is described by the {\em emissivity} coefficient, indicated by $\epsilon$. 
For pyroclastic particles at FLIR wavelengths (8--14 $\mu$m), if the mixture is a trachyte/rhyolite,
then an emissivity of 0.975 is appropriate \citep{Harris2013}. Thus to convert between spectral
radiance and kinetic temperature we use:

\begin{equation}\label{eq:blambda}
B_{\lambda}(T) = \epsilon \, \frac{2 h c^2}{\lambda^5} \frac{1}{e^{\frac{hc}{\lambda k_BT}}-1}
\end{equation}
where $h$ is the Planck's constant, $c$ is the speed of light, $k_B$ is the Boltzmann's constant and
$T$ is the temperature.

\subsection{Absorption by atmospheric and volcanic gases}
Thermal cameras installed to video volcanic plumes are typically insatalled at distances of several kilometers from the source. 
This allows safe measurements, and the full ascent history from vent to point of stagnation to be imaged.  
Over such distances, the effect of atmospheric absorption will be non-negligible \citep{Harris2013}. 
This effect becomes more important as humidity increases, because water droplets have high absorption properties at TIR wavelengths. 
Volcanic gases also have a significant effect on absorbing emitted radiation in the TIR \citep{sawyer2006}. 
Therefore, to apply Eq.\eqref{eq:Ilambda} the absorption coefficients of the atmospheric and volcanic gases need be taken into account.

Absorption by gases can be computed using Eq.\eqref{eq:differentialSch}, so that the resulting coefficient is the sum of the coefficient of
the $N_{ph}$ phases, whereas the emissivity is a weighted average:
\begin{equation*}
K_{mix} = \sum_j^{N_{ph}} K_j; \qquad
\epsilon_{mix} = \frac{\displaystyle{\sum_j^{N_{ph}}} K_j \epsilon_j}{K_{mix}}
\end{equation*}
Analogously to the expression of $K_s$ for particles, the absorption coefficient for gases can also
be expressed as the product of the specific absorption coefficient $A_j$ (which depends only on gas material properties) times the gas
bulk density $K_j(\mathbf{x}) = A_j \, \rho_j(\mathbf{x})$.

For example, in volcanic ash plumes we may want to consider the presence of water vapor, carbon and sulphur dioxide. In such a case, at any point $\mathbf{x}$:

\begin{equation}\label{eq:kmix}
K_{mix} = \sum_j^{N_{ph}} K_j = \sum_j^{N_{ph}} A_j \,\rho_j(\mathbf{x}); \qquad
j=s,H_2O,CO_2,SO_2,Air
\end{equation}

For typical eruptive conditions (involving water vapour as the main volcanic gas), $A_s$ and $A_w$ are of the same order of magnitude.
Fig.~\ref{fig:waterAbC} shows the absorption coefficient for atmospheric water vapor at sea level. 
The specific absorption coefficient is obtained by dividing by the corresponding density, obtaining $A_w \simeq 1$ in the waveband 8--14 $\mu$m.
It is important to notice that, whereas specific absorption coefficients for particles are almost independent from the wavelength (under the assumptions of large particles) for gases this dependency can be significant. 
We therefore consider that the coefficients $A_j$ for gases represent an average value over a relatively narrow band, corresponding to the wavelength window of TIR detectors (8--14 $\mu$m). 
%
\subsection{Atmospheric background radiation}
The background atmospheric radiation (the first term of Eq.\ref{eq:Ilambda}) also contributes to
the detected TIR radiance.
Whereas the center of an ash plume is generally opaque to transmission of background thermal radiation (meaning that this term is negligible along an optical path crossing the axis of an ash plume), part of the background atmospheric radiation can be trasmitted through a gaseous plume or through the diffuse margins of an ash-laden plume, where particle concentration is much lower.
The treatment of background radiation begins with an estimate of the spectral radiance in the absence of the plume at a distance $L$ from the source, $L$ being larger than the distance of the observer from the plume axis (see Fig.~\ref{fig:sketch}).
We will show in Section \ref{inversion} how this can be done in practical cases.\\

In summary, the at-detector spectral radiance $I_{\lambda}(L)$ at wavelength $\lambda$ associated to emission/absorption balance from a gas--particle mixture in the atmosphere can be computed using an electromagnetic model by specifying the following variables and parameters along each optical path received by a detector:
\begin{itemize}
\item the Sauter diameter $d_s$ of the particle distribution (Eq. \ref{eq:k});
\item the spatial distribution of particle volumetric concentration $\varepsilon$ (Eq. \ref{eq:k});
\item the spatial distribution of temperature $T$ (Eq. \ref{eq:blambda});
\item the material (either gas or solid) emissivity $\epsilon$ (Eq. \ref{eq:blambda});
\item the specific absorption coefficients $A_j$ for each gas species (Eq. \ref{eq:kmix});
\item the bulk density distribution $\rho_j$ of each gas species (Eq. \ref{eq:kmix}).
\end{itemize}
Whereas material properties (emissivity, specific absorption coefficients) can be obtained from laboratory measurements, the spatial distribution of gas and particles, plus their variation in density and temperature need to be derived from a fluid-dynamic model that describes the dynamics of the volcanic plume for specific vent conditions.

\section{Plume model}
\label{model}
In this section, we present a one-dimensional fluid-dynamic model that can be applied reconstruct the particle concentration and temperature distribution in the volcanic plume to allow computation of the IR radiation field. 
In Section \ref{inversion} we will adopt this model to solve the inverse problem.

Our model is based upon the \citet{woods1988} model, which describes the transport of volcanic ash and gases in the atmosphere assuming a {\em dusty gas} approximation \citep{carrier1958,marble1970}.
Accordingly, full local kinetic and thermal equilibrium between gas and particles are assumed, i.e. all components of the eruptive mixture are assumed to have the same velocity and temperature fields.
In this work we will assume that the erupting mixture is composed of volcanic gas (subscript $e$), solid particles (subscript $s$) and atmospheric gas (subscript $\alpha$). 
We denote the dusty-gas mixture with subscript $\beta$. Thermodynamic properties of the dusty gas are reported in \ref{thermo}.

Woods' transport model can be derived formally from the full three-dimensional dusty-gas model under the following assumptions \citep{morton1956, morton1959, Wilson1976, list1982, papanicolaou1988, woods1988, fannelop2003, kaminski2005, ishimine2006, plourde2008}:
\begin{itemize}
 \item The Reynolds number is large and turbulence is fully developed, so that thermal conduction and viscous dissipation can be neglected.
 \item Pressure is constant across a horizontal section and is adjusted to atmospheric pressure at the vent (i.e., the model cannot be applied to underexpanded jets).
 \item The plume is {\em self-similar}, i.e., the time-averaged profiles of vertical velocity, density and temperature across horizontal sections at all heights have the same functional form (either {\em top-hat} or Gaussian).
 \item The time-averaged velocity field outside and near the plume edge is horizontal.
 \item Stationary input conditions and steady atmosphere (no wind).
 \item Radial symmetry around the source.
\end{itemize}

We seek an axisymmetric solution in a poloidal plane $(r,z)$ (with unit vectors $(\hat{r},\hat{z})$) of the dusty gas conservation equations with the following form:
\begin{align}
 \rho(r,z) & = \begin{cases} \beta(z)\,, & \mbox{if } 0\leq r < b(z)\\ \alpha(z)\,, & \mbox{if } r
\geq b(z) \end{cases}\label{eq:rhoS}\\
 u(r,z) & = \begin{cases} +U(z)\hat{z}\,, & \mbox{if } 0\leq r < b(z)\\ -\Ue(z) \hat{r}\,, &
\mbox{if } r = b(z)\\ -u_\epsilon(r,z) \hat{r}\,, & \mbox{if } r > b(z)\\ u_\epsilon \to 0 &
\mbox{if } r \gg b(z) \end{cases}\label{eq:uS}\\
 p(r,z) & = p(z)\label{eq:pS}\\
 T(r,z) & = \begin{cases} \Tb(z)\,, & \mbox{if } 0\leq r < b(z)\\ \Ta(z)\,, & \mbox{if } r \geq b(z)
\label{eq:TS}\end{cases}
\end{align}
where $\rho$, $u$, $p$ are the flow density, velocity and pressure, $b$ is the plume radius, $U$ is the plume axial velocity and $U_{\epsilon}$ is the entrainment velocity.
Here  we assume a ``top hat'' self-similar profile (i.e., a constant profile) for all variables, so as to simplify computation with a view to efficiently solve the inverse problem. 
By imposing the top-hat plume solution, the resulting mass, momentum and energy balance equations read \citep{woods1988}:
%
\begin{equation}\label{eq:Woods}
 \begin{sistema}
  \de_z (\rho_jU b^2) = 0\quad (j=s,e)\vspace{2pt}\\
  \de_z (\beta U b^2) = 2 \alpha b \Ue \vspace{2pt}\\
  \de_z (\beta U^2 b^2) = (\alpha - \beta)g b^2 \vspace{2pt}\\
  \de_z \left( \beta U b^2\,\Cb\Tb \right) = \left(\Ca \Ta\right) \de_z(\beta U b^2) + \frac{U^2}{2}
\de_z(\beta U b^2) - g\alpha U b^2\,.
 \end{sistema}
\end{equation}

The first equations state that the mass fluxes of volcanic gases and particles must be conserved, so that their value is constant along the plume axis ($\rho_j(z)$ is their bulk density). $g$ is the gravity acceleration.
The remaining unknowns are $\beta(z)$, $U(z)$, $b(z)$ and $\Tb(z)$.
The system is closed by opportune equations of state expressing the mixture density as a function of temperature $\beta=\beta(T_{\beta})$ (thermal equation of state) and specific heats $C_{\beta}$ and $C_{\alpha}$ (caloric equation of state). For a dusty gas, thermodynamic properties are computed locally from the properties of each component of the mixture. 
Thermodynamic closure equations are reported in \ref{thermo}.

Finally, the ambient density $\alpha(z)$, the ambient temperature $\Ta(z)$ and the dependence of $\Ue$ on other unknowns (the entrainment model) must be given.
For the atmosphere, we assume a linear temperature decrease at a constant rate in the troposphere (as is appropriate for weak plumes) and a hydrostatic density profile.
The entrainment velocity $\Ue$ is expressed as  $\Ue = k U \eta\left(\beta/\alpha\right)$, where $k$ is a dimensionless entrainment coefficient and $\eta$ is an arbitrary function of the density ratio \citep{fannelop2003}. 
When $\eta = 1$ we recover the model of \citet{morton1956}, and if $\eta(x) = \sqrt{x}$ we obtain the model of \citet{ricou1961}. 
In this work, we will always assume $\eta = 1$.

Following \citet{fannelop2003}, we also define some new compound variables :
\begin{eqnarray}
\label{eq:qmf}
Q_j &=& \rho_jU b^2 \quad (j=s,e) \label{eq:Qj}\\
Q &=& \beta U b^2 \label{eq:Q}\\
M &=& \beta U^2 b^2 \label{eq:M}\\
F &=& \left[(\alpha - \beta) + \alpha \frac{\chi Q_m}{Q - Q_m}\right]U b^2 \label{eq:F}
\end{eqnarray}
This allows the transport equations to be expressed in a more compact form.

In Eq.~\eqref{eq:F}, $Q_m = Q_s - (\psi_e - 1)Q_e$ and $\chi = \frac{(\chi_s-1) Q_s + (\chi_e-1)Q_e}{Q_m}$. Note that $Q_m=Q_s$ if the emitted gas is identical to the atmospheric gas ($\psi_e = R_e/R_{\alpha}=1$) whereas $\chi$ accounts for the difference in the caloric properties of the erupted mixture with respect to the atmosphere.

The expression for $F$ (Eq.~\ref{eq:F}) represents a modification of the buoyancy flux for a dusty-gas plume. It takes the classical form  $F=(\alpha - \beta) U b^2$ \citep{fannelop2003} for $Q_m=0$, i.e. for a single-component gas plume.
It is also convenient to introduce the non-dimensional variables $\zeta=z/L$, $q=Q/Q_0$, $m=M/M_0$ and $f=F/F_0$, where $L=\frac{Q_0}{\sqrt{\alpha_0 M_0}}$ and the subscript $0$ indicates values at $\zeta=0$. We also define $q_m = Q_m/Q_0$ and $\phi=F_0/Q_0$.\\

The general form of equation \eqref{eq:Woods} in terms of the new variables ($q,m,f$) can be solved numerically as a function of the vent conditions, namely the vent radius, the mass flow rate of all gaseous and solid components and its temperature \citep{cerminara2014}. By transforming back into the dimensioned variables (algebraic transformations are reported in \ref{nondim}), numerical results provide, at each height, the mixture bulk density $\beta(z)$, vertical velocity $U(z)$, temperature $T_{\beta}(z)$ and plume radius $b(z)$. The bulk density of all dusty-gas components $\rho_j(z)$ can then be computed, and the three-dimensional distribution of all variables can be recovered from the one-dimensional results by applying the self-similarity hypothesis (Eqs.\ref{eq:rhoS}--\ref{eq:TS}).

\subsection{Approximate solution of the plume model}
Instead of solving the complete model of Eq.~\eqref{eq:Woods}, in this work we adopt a simplified plume model. This further decreases computation time. This approximation asymptotically describes the dusty-plume dynamics within the Boussinesq limit (i.e., for small density contrasts) and for a non-stratified atmosphere. This regime is similar to that explored by \citet{morton1956} for
single-phase gas plumes.
Such an approximation holds for an intermediate region ``far enough'' from the vent and ``well below'' the maximum plume height (where neutral buoyancy conditions are reached). These conditions are met for $q \gg q_m$ and $q \gg f$.
Because, in this regime, the modified buoyancy flux is constant $f=1$, the second condition reduces to
$q \gg 1$, which is satisfied far enough from the vent (after sufficient air has been entrained). To
be more quantitative, $4b(0) < z < H$ where $H=\frac{\gamma R T}{g}\approx 10$ km is the typical
length scale for atmospheric stratification.

In this approximation, the system of equations \eqref{eq:Woods} reduces to \citep{cerminara2014}:
\begin{equation}\label{eq:QMF}
 \begin{sistema}
 \displaystyle
  \de_{\zeta} (q) = v_q \sqrt{m} \vspace{2pt}\\
   \displaystyle \de_{\zeta} (m) = v_m \frac{q}{m} \vspace{2pt}\\
  \de_{\zeta} (f) = 0 \vspace{2pt}\\
 \end{sistema}
\end{equation}
where we have introduced the following parameters:
\begin{align}
\label{eq:vq}
 v_q & = 2k\\
\label{eq:vm}
 v_m & = \frac{g F_0 Q_0^2}{\alpha_0^{1/2} M_0^{5/2}}\,(1-\gamma)\\
 \gamma &  = \frac{(\chi +1 )q_m}{\phi}
\end{align}

This system is formally equivalent to the \citet{morton1956} equations but with different
dimensional constants $v_q$ and $v_m$. These now account for the multiphase nature of the erupting
mixture. Morton's equations are recovered by setting $\gamma=0$ and $\chi=0$ (single
phase plume).

Eqs.~\eqref{eq:QMF} have an asymptotic analytical solution expressed by:
\begin{equation}
\label{eq:analytical}
\begin{sistema}
q(\zeta) = \sqrt{ \frac{4 v_q}{5 v_m} \left[ \left( \frac{3}{4} \sqrt{\frac{4 v_q v_m }{5}} \zeta +
1  \right)^{10/3} + \frac{5 v_m}{4 v_q} - 1\right]}\\
m(\zeta) = \left[ \frac{3}{4} \sqrt{\frac{4 v_q v_m }{5}} \zeta + 1 \right]^{4/3}\\
f(\zeta) = 1
\end{sistema}
\end{equation}
which also has the correct asymptotic behaviour for $\zeta \gg 1$.
This solution can be viewed as a function of the boundary values of the flow variables and the model
parameters, namely $(v_q,v_m,L,\phi,\chi,q_m)$.

\begin{figure}
\centering
\subfloat[][]{\includegraphics[width=0.4 \columnwidth]{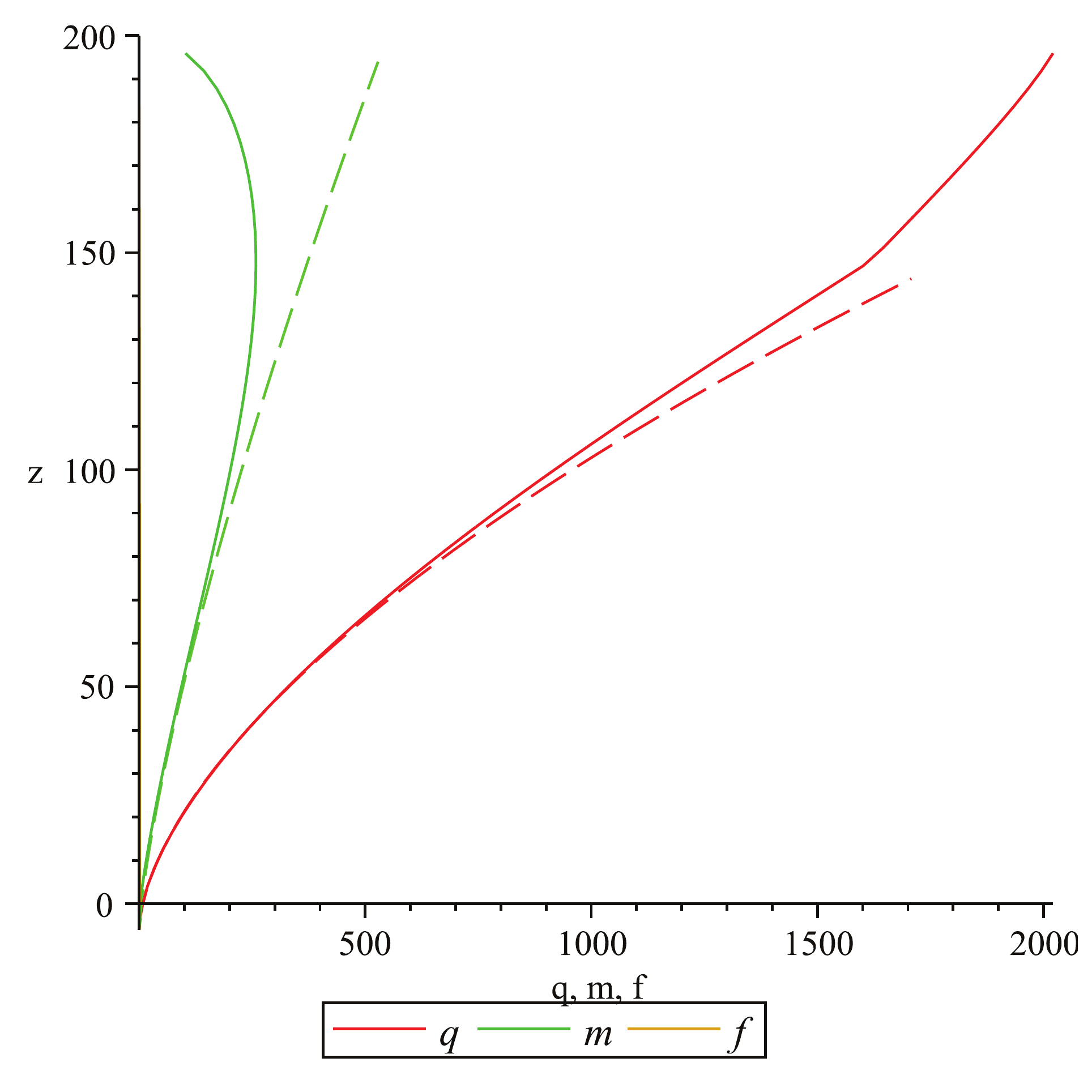}}
\subfloat[][]{\includegraphics[width=0.4 \columnwidth]{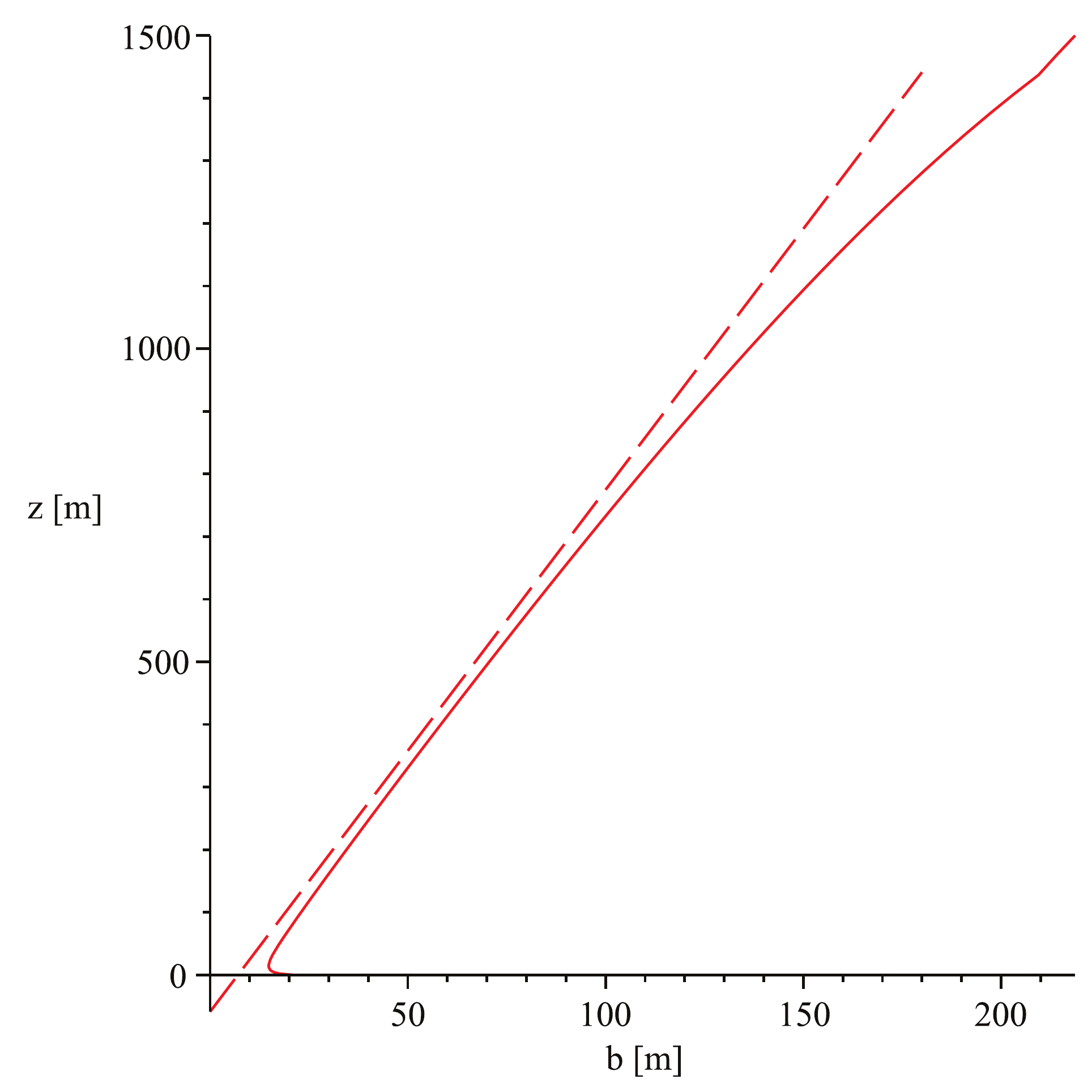}}\\
\subfloat[][]{\includegraphics[width=0.4 \columnwidth]{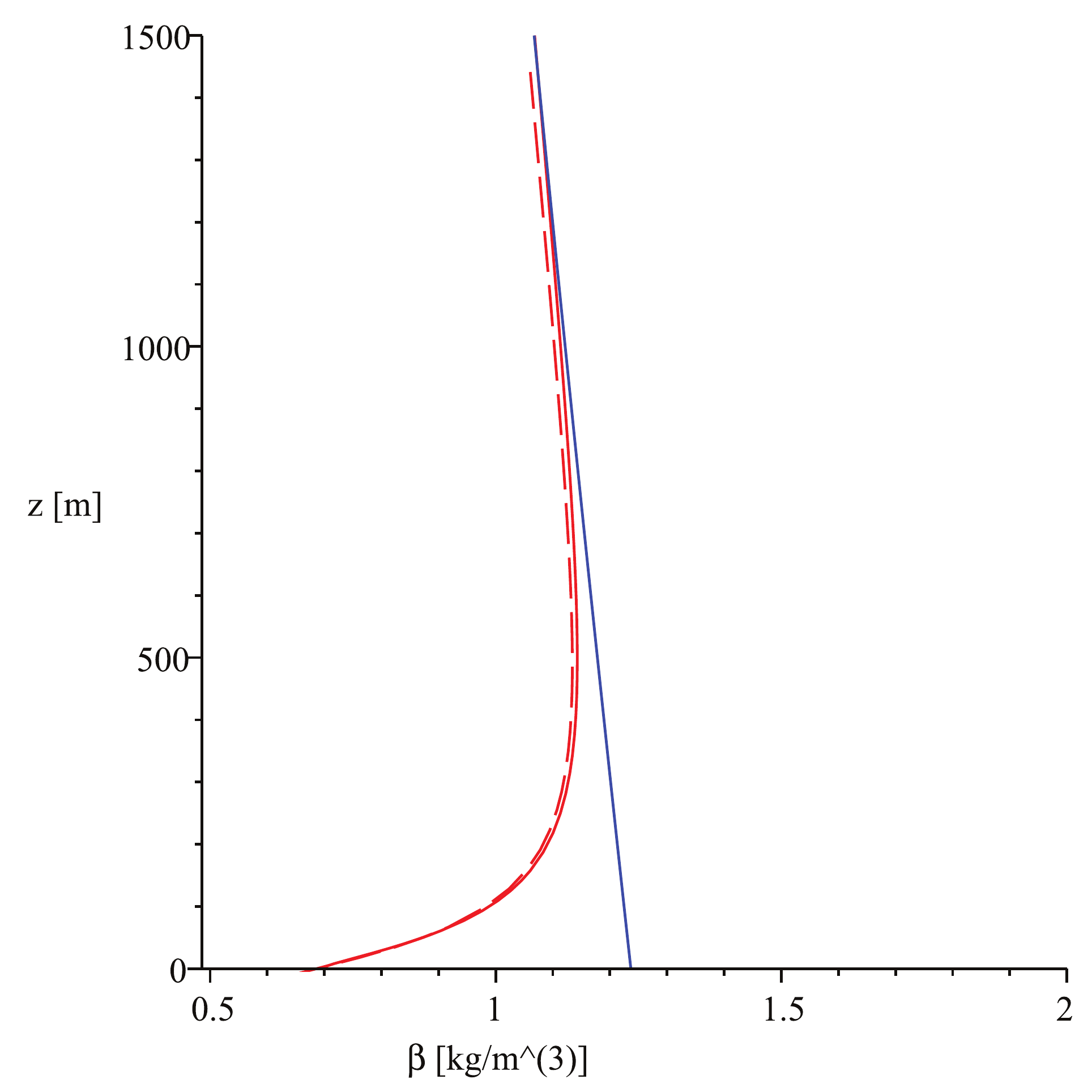}}
\subfloat[][]{\includegraphics[width=0.4 \columnwidth]{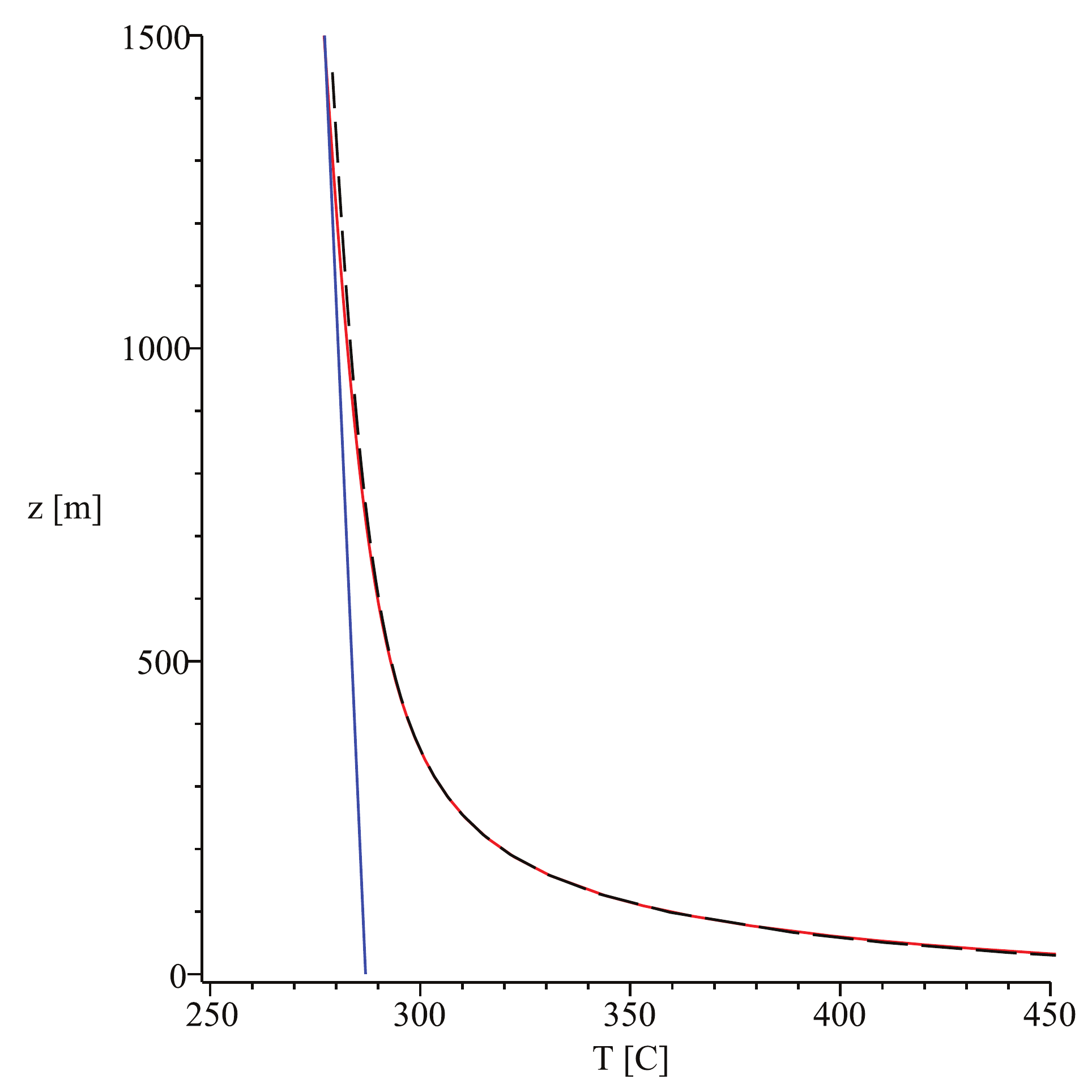}}
\caption{Comparison between the numerical solution of the \citet{woods1988} model (solid lines) and
the approximated analytical solution (Eq. \ref{eq:analytical}, dashed lines). a) Non-dimensional
variables $(q,m,f)$ as a function of the non-dimensional length scale $\zeta$; b) dimensioned plume
radius $b(z)$; c) plume density $\beta(z)$ (Eq.~\ref{eq:betadiq}); d) plume temperature
$T_{\beta}(z)$ (Eq.~\ref{eq:tbetadiq}). The straight solid lines represent the atmospheric
profiles.}
\label{fig:comparison}
\end{figure}

In Figure \ref{fig:comparison} we compare the numerical solution of Eq.~\ref{eq:Woods} with the analytical solution of Eq.~\ref{eq:analytical}, assuming vent conditions for a typical weak plume at Santiaguito \citep{sahetapy2009,valade2014}, as reported in Tab. \ref{tab:ventc} and discussed in Sect.~\ref{inversion}. Consistently with this approximation, the analytical solution closely fit the complete model solution in the central part of the plume. However, it differs near $z=0$ and near the plume top. Note that the density and temperature profiles (plots c) and d) have been derived from the non-dimensional variables $(q,m,f)$ by using the stratified atmospheric profiles $\alpha(z)$ and $T_{\alpha}(z)$ into Eqs.~\eqref{eq:betadiq} and \eqref{eq:tbetadiq}. 
In this way we have taken into account atmospheric stratification while maintaining the simple structure of the equations of Morton's approach (which holds in a non-stratified environment).

\begin{table}
\centering
\begin{tabular}{cc}
\hline
$U_0$ & 5 m/s \\
$T_0$ & 578 \celsius \\
$b_0$ & 21 m\\
$Q_e$ & 571 kg/s\\
$Q_s$ & 414 kg/s\\
$\frac{{\rm d}T_{\alpha}}{\rm dz}$ & - 4.4 \celsius/km\\
$T_{\alpha 0}$ & 15 \celsius\\ 
$k$ & 0.1 \\
\hline
$\beta_0$ & 0.447 kg/m$^3$\\
$n$ & 0.58 \\
$Q$ & 985 kg/s \\\hline
\end{tabular}
\caption{Initial and boundary conditions of weak plume simulation.\label{tab:ventc}}
\end{table}

\section{Coupled forward model}
\label{fluid-em}
\begin{figure}
\centering
\includegraphics[width=0.6\columnwidth]{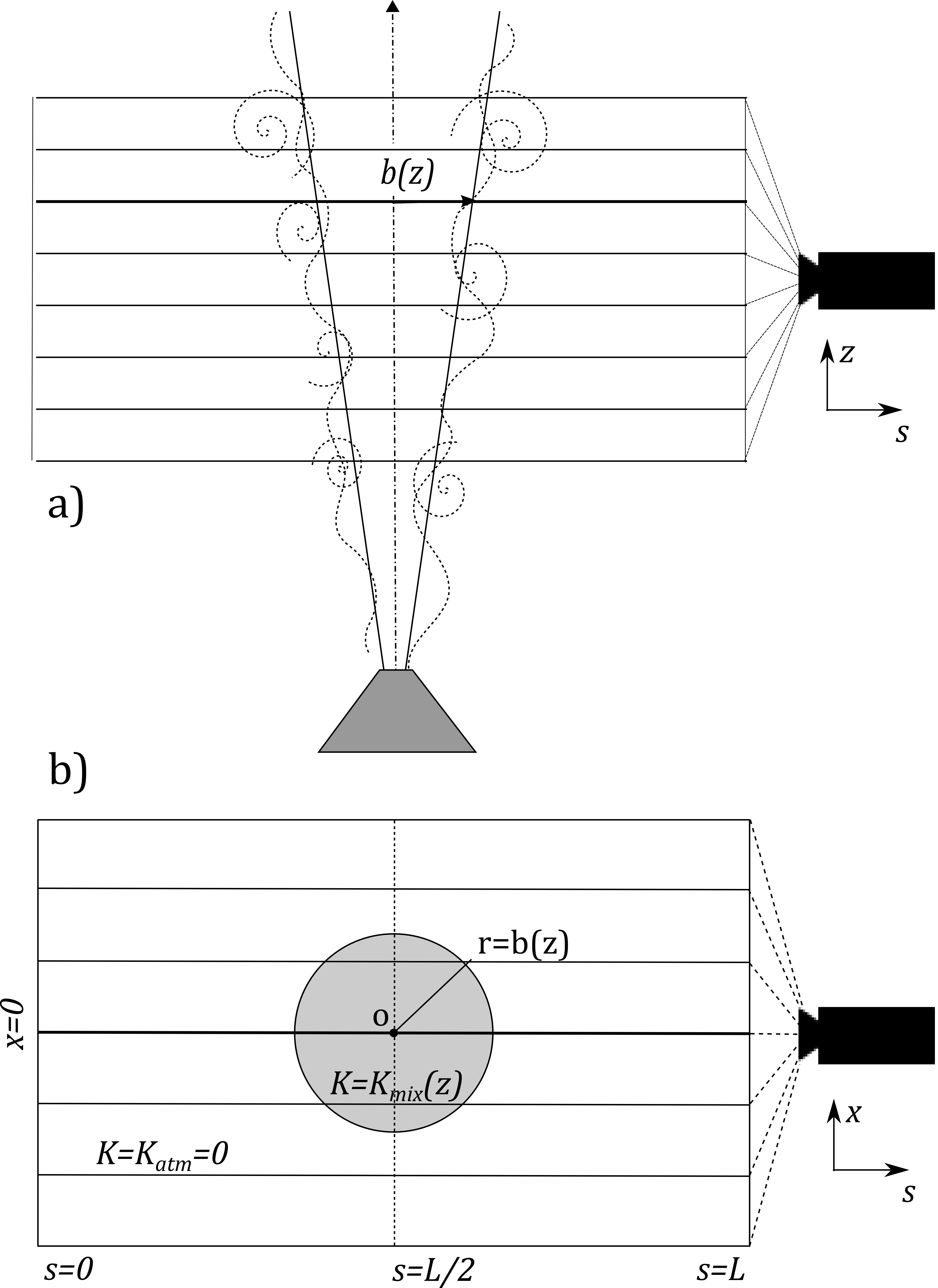}
\caption{Geometric configuration adopted for the calculation of the IR intensity. a) Side view. b)
Top view of a cutting plane orthogonal to the plume axis (point {\bf o}). The plume radius is
represented by the gray-shaded region (top-hat approximation).}
\label{fig:sketch}
\end{figure}

We assume that the bulk density $\rho_j$ of each gaseous and solid component is given at every point of the domain (Figure \ref{fig:sketch}). For each component, the specific absorption coefficients $A_j$ and emissivity $\epsilon_j$ (which depend only on the material) are assumed to be known. Now,
\begin{enumerate}
\item The absorption coefficient of the mixture $K_\textup{mix}$ can be estimated at any point by using
Eq.~\ref{eq:kmix}.
\item Along every ray in Figure \ref{fig:sketch} we can compute the optical thickness $\tau(s)$
(Eq.~\ref{eq:taus}) by integrating $K_\textup{mix}$ along the ray trajectory (which is assumed to be
a straight line).
\item We can compute the Planck function of the mixture (Eq.~\ref{eq:blambda}) at each point of the
domain as a function of the local temperature and the local mixture emissivity.
\item Finally, the background radiation $I_0$ is estimated at some point behind the plume (e.g., at
$s=0$ in Fig. \ref{fig:sketch}), taken as the image horizon.
\end{enumerate}
With these ingredients, we can compute the radiation intensity along a discrete number of rays forming the electromagnetic image of the domain $\Omega$.
It is worth noting that, usually, commercial devices give the temperature as output image, not the intensity. To derive the temperature image from the TIR intensity, the Planck function has to be inverted.
\subsubsection{Geometric approximations}
To simplify the problem, we have adopted the following geometric approximations:
\begin{itemize}
\item the camera is far enough from the plume so that rays can be considered as parallel;
\item rays are assumed to cross the plume axis orthogonally;
\item effect of plume bending (eventually due to wind) are corrected by means of image processing
techniques.
\end{itemize}

With these hypotheses, the geometric configuration required to construct the IR image is sketched in Fig.~\ref{fig:sketch}b, where the plume axis is oriented normally to the image plane at $r=0$. Radius $b(z)$ depends on the height above the vent and the concentration and temperature fields are constant inside the circle and zero outside.

By adopting a top-hat assumption for the plume profile, the radiant intensity can be computed analytically under the further simplification that the emission/absorption of the atmosphere can be neglected (this is reasonable if the distance of the camera from the plume is not too large, indicatively less than about 10 km).
In this case, the absorption coefficient is taken equal to zero outside the plume, whereas the value of $K_\textup{mix}(z)$ within the plume can be computed starting from the analytical solution of the plume model, by expressing the mixture density $\beta$ in terms of the non-dimensional variable $q$
(Eqs. \ref{eq:betadiq} and \ref{eq:analytical})
\begin{equation}
\label{eq:kmixdiq}
K_\textup{mix}(z) = A_s \rho_s + A_e \rho_e = \left( A_s q_s + A_e q_e \right) \frac{\beta}{q} =
\left( A_s q_s + A_e q_e \right)\alpha \frac{(q + \chi q_m)}{(\phi+q)(q-q_m)}
\end{equation}
with $K_\textup{mix}(z)$ depending on $z$ only through $q=q(z)$. We also define the specific absorption coefficient of the mixture
\begin{equation}
\label{eq:Am}
A_m=\left( A_s q_s + A_e q_e \right)
\end{equation}
so that $K_\textup{mix} = A_m\,\beta/q$ and $A_m$ is an initial mixture parameter that does not
depends on the position along the plume.

With reference to Fig. \ref{fig:sketch}b), along each ray we identify the points $s_1$ and $s_2$
where the ray crosses the edge of the plume. For $-b<x<b$ their coordinates are $s_1=L/2 -
\sqrt{b^2-x^2}$ and $s_2=L/2 + \sqrt{b^2-x^2}$ and the optical thickness is then simply 
\begin{equation}
\tau(s,z) =
\begin{sistema}
0 \qquad \qquad \qquad \qquad 0<s<s_1\\
K_\textup{mix}(z) \, (s-s_1) \qquad s_1<s<s_2\\
K_\textup{mix}(z) \, (s_2-s_1) \qquad s_2<s<L\\
\end{sistema}
\end{equation}

\begin{figure}
\centering
\includegraphics[width=0.5\columnwidth]{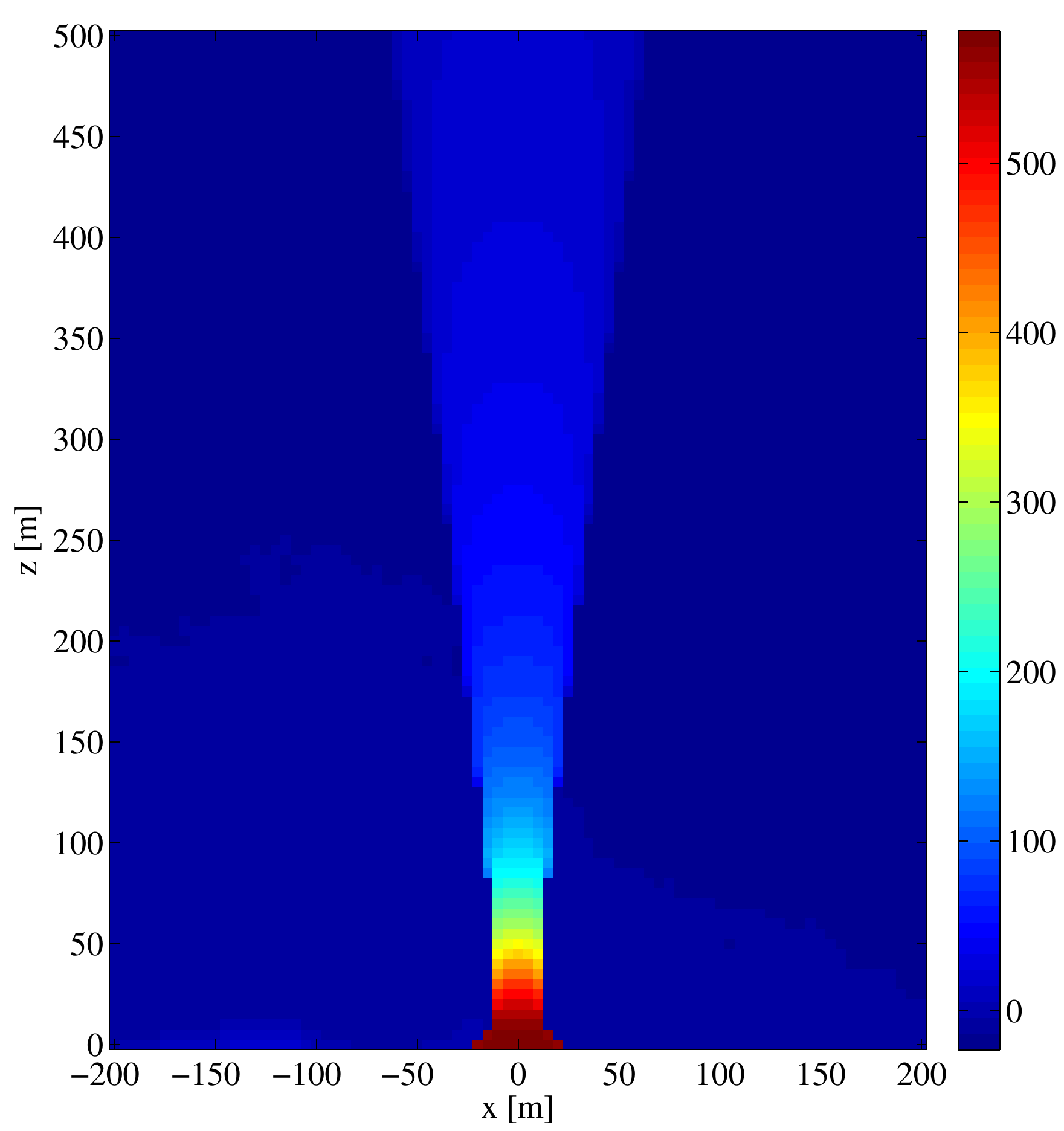}
\caption{Synthetic IR image obtained from the 1D approximated model (Eq.~\eqref{eq:analytical}). The
electromagnetic intensity $I_L$ (Eq.~\ref{eq:I_L}) has been converted into a temperature image by
means of the Planck's equation \eqref{eq:blambda}.}
\label{fig:syntheticIR}
\end{figure}

Because in this example we have assumed that the optical thickness of the atmosphere is zero, we find
that 
\[
\tau_L(z) = K_\textup{mix}(z) \, (s_2-s_1) = 2 K_\textup{mix}(z) \sqrt{b^2-x^2} \theta(b^2-x^2)
\]
where $\theta$ is the Heaviside step function.
To compute the Planck constant, we also neglect possible variations in emissivity along the optical path and assume that $\epsilon = 1$ across the plume. This is equivalent of assuming that the ash behaves as a black body \citep{Harris2013}.
With these hypotheses, the Planck function depends only on the vertical coordinate $z$ through $T_{\beta}(z)$ (Eqs. \ref{eq:tbetadiq} and \ref{eq:blambda}).
By solving the integral, we obtain:
\begin{equation}
\label{eq:I_L}
I_L(x,z) = I_0 e^{-\tau_L} + B(z)\left( 1 - e^{-\tau_L}\right)
\end{equation}
In Figure \ref{fig:syntheticIR} the synthetic TIR image of the plume given in Figure \ref{fig:comparison} is shown ($I_0$ has been computed by assuming a uniform background temperature of 288K). 
The thermal image obtained from the approximated model does not appreciably differ from that obtained from the Woods' model (not shown), because the profiles of radius, density and temperature are almost coincident (see Fig.~\ref{fig:comparison}b-d).

\section{Inverse model and application}
\label{inversion}
The coupled fluid--electromagnetic model described in the previous sections provides a synthetic infrared image of a gas--particle plume, that we have called $I_L(x,z)$. This is a complex, non-linear function of the flow conditions at the vent and of the material properties of volcanic gases and particles and of the atmosphere. More specifically, assuming that the material properties are known and neglecting the emission/absorption contribution of the atmosphere, the synthetic image can be expressed as a function of the plume model boundary values and parameters and of the specific absorption coefficient of the mixture $A_m$ (given by Eq. \ref{eq:Am}):
\begin{equation}
\label{eq:unknowns}
I_L = I_L(v_q,v_m,L,\phi,\chi,q_m,A_m)
\end{equation}
Using the algebraic transformations of \ref{parinv} we can express $I_L$ as a function of $(b_0, U_0, T_0, n_0, k, d_s)$ where $b_0$, $U_0$, $T_0$, $n_0$ are the plume radius, velocity, temperature and gas mass fraction at $z=0$,  $k$ is the air entrainment coefficient and $d_s$ is the equivalent Sauter diameter of the grain size distribution. 
Note that $z=0$ may not correspond to the vent quota but instead to the minimum height of the acquired image. 
It is also worth recalling here that we assume that $d_s$ does not change during plume development, i.e., that particle grain size distribution does not change in the plume. However, if this is not the case \citep[e.g., because of the effect of particle loss from the plume margins -- ][]{woods1991} the procedure can still be applied but the plume solution (Eq.~\ref{eq:analytical}) and the TIR radiation (Eq.~\ref{eq:I_L}) should be modified accordingly.

This synthetic image can now be compared to the actual TIR images captured during the volcanic event. 
We will demonstrate in this section how it is possible to estimate the parameters in Eq.~\ref{eq:unknowns} by means of inversion procedures.
To do this, TIR images must be preliminary processed in order to obtain an average experimental intensity image $I_E(x,z)$ and a background image $I_0(x,z)$. The minimum of the difference $||I_E-I_L||=f(v_q,v_m,L,\phi,\chi,q_m,A_m)$ is then sought in the parameter space to find the
eruptive conditions which best fit the observation.
\subsection{Image processing}
TIR video used here provide a sequence of $N+1$ IR images $P_i (i=0,...,N)$ of a developing plume, acquired at a fixed time rate. Usually, commercial devices automatically convert the digital intensity image registered by the charge coupled device (CCD) into a 8 or 16 bit temperature image. 
We will here assume that the first image $P_0$ represents the time immediately before the eruption and that $P_1$ is the first image of the erupting plume. Because some time is needed for the plume to develop, we will also assume that the flow can be considered stationary between frames $P_m$ and $P_f$.
Under such assumptions, we can compute an average TIR image $P_a = \frac{\sum_{i=m}^f P_i}{f-m+1}$.

By means of image processing techniques \citep{valade2014} the plume trajectory is extracted from $P_a$ and the region of interest along the axis is selected. If the plume axis is bent (as a result of wind or source anisotropy) the images $P_0$ and $P_a$ are corrected by means of geometric transformations (rotation and dilatation). 
This is also used to correct possible image distortions associated to camera orientation.

Finally, Eq.~\eqref{eq:blambda} is applied to thermal images $P_a$ and $P_0$ to obtain the
experimental intensity image $I_E(x,z)$ and the atmospheric background $I_0(x,z)$, where $z$ run
along the axial direction and $x$ along the horizontal direction perpendicular to the camera optical
axis.
\subsection{TIR dataset for an ashy plume at Santiaguito}
We use a set of TIR images of an explosive ash emission that occured at Santiaguito volcano (Guatemala) in 2005 \citep{sahetapy2009}. The duration of ash emission is about $\Delta t \simeq 300$ s, and was sampled at 30 Hz.

\begin{figure}
\centering
\subfloat[][]{\includegraphics[width=0.5 \columnwidth]{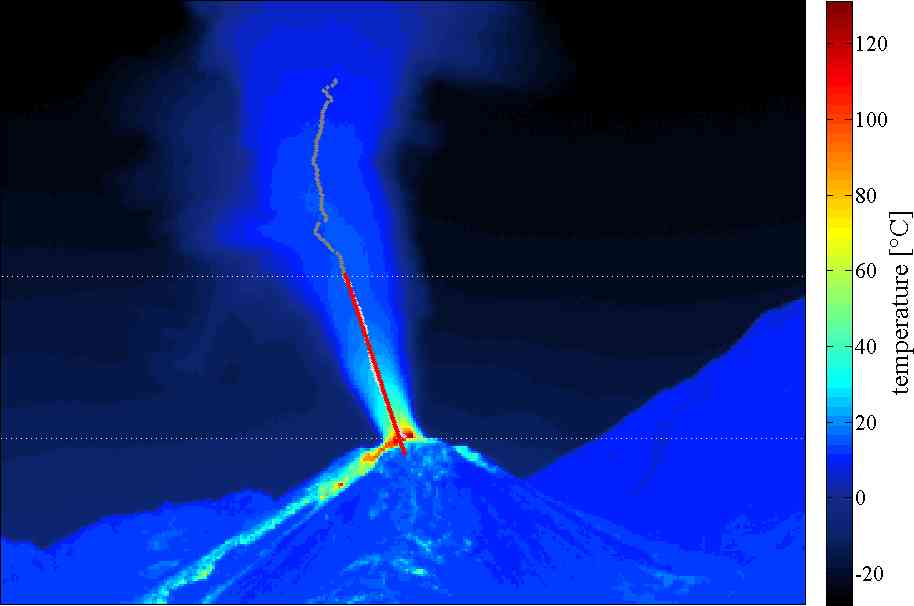}}\quad
\subfloat[][]{\includegraphics[width=0.4 \columnwidth]{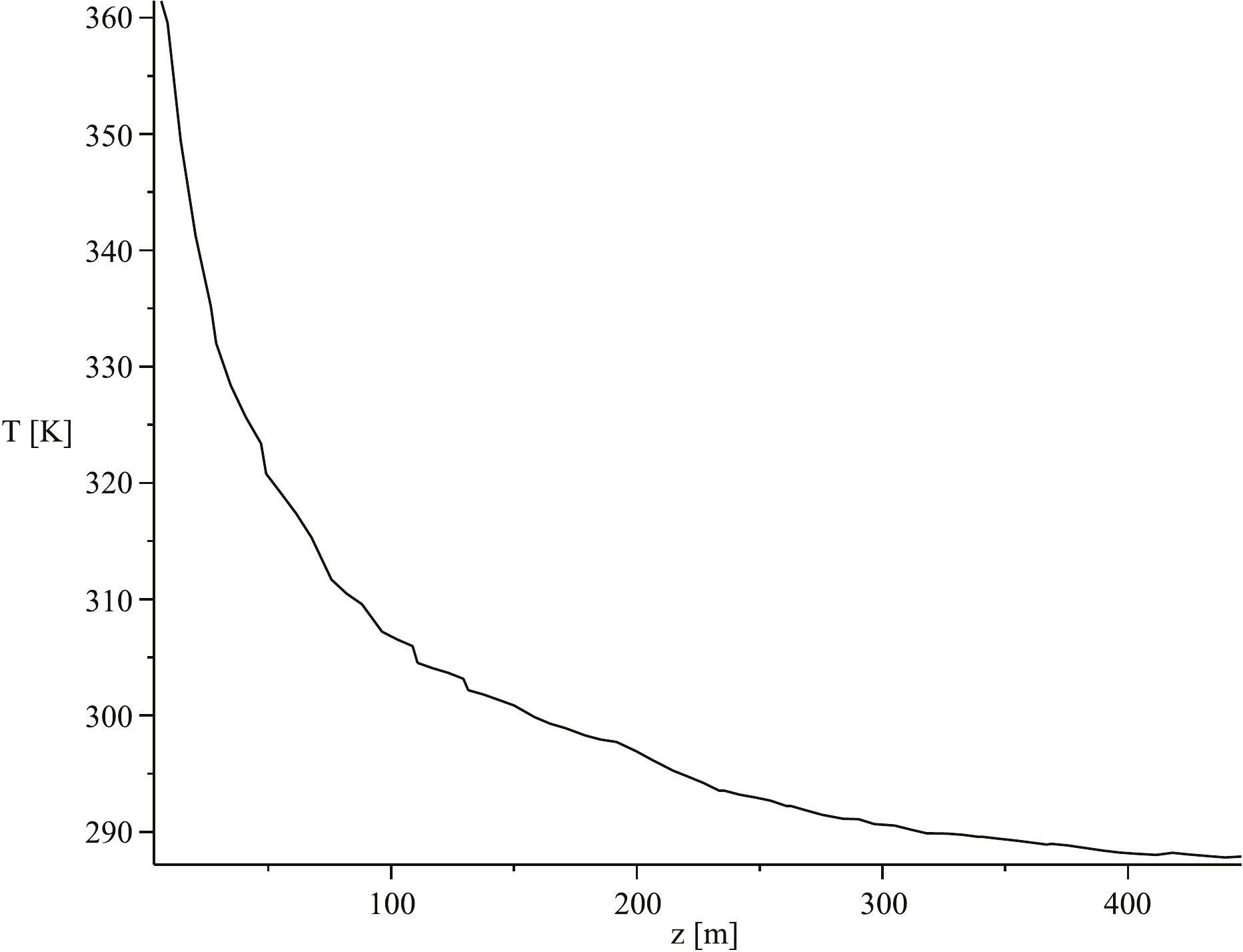}}
\caption{a) Averaged image computed from a set of thermal infrared images recorded with a FLIR
camera, imaging the stationary emission of a sustained volcanic ash plume at Santiaguito volcano
(Guatemala). The red dots represent the extracted plume axis \citep{valade2014}. b) Temperature
values along the plume axis.}
\label{fig:meanlmg}
\end{figure}

\begin{figure}
\centering
\subfloat[][]{\includegraphics[width=0.45\columnwidth]{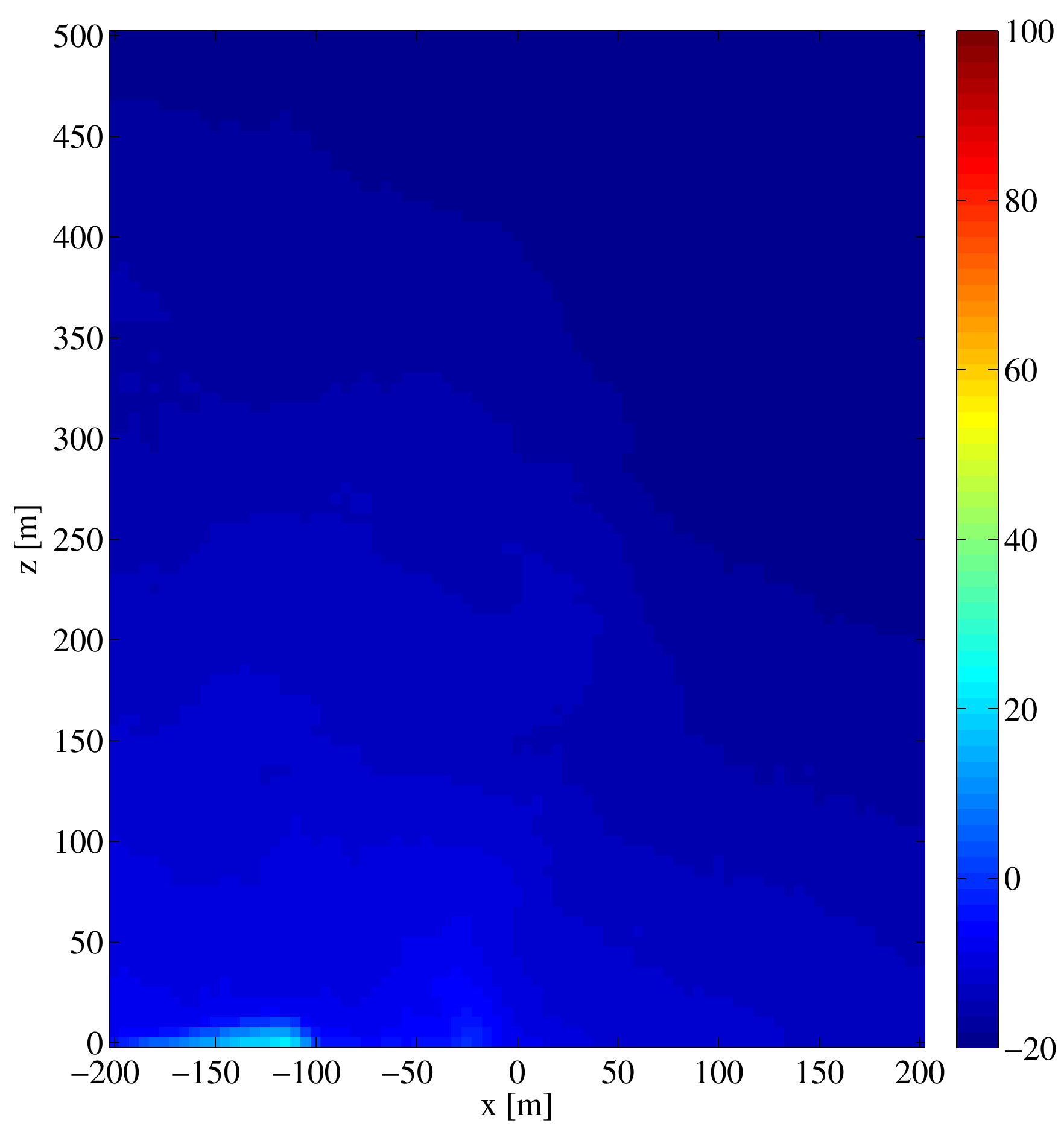}}\quad
\subfloat[][]{\includegraphics[width=0.45\columnwidth]{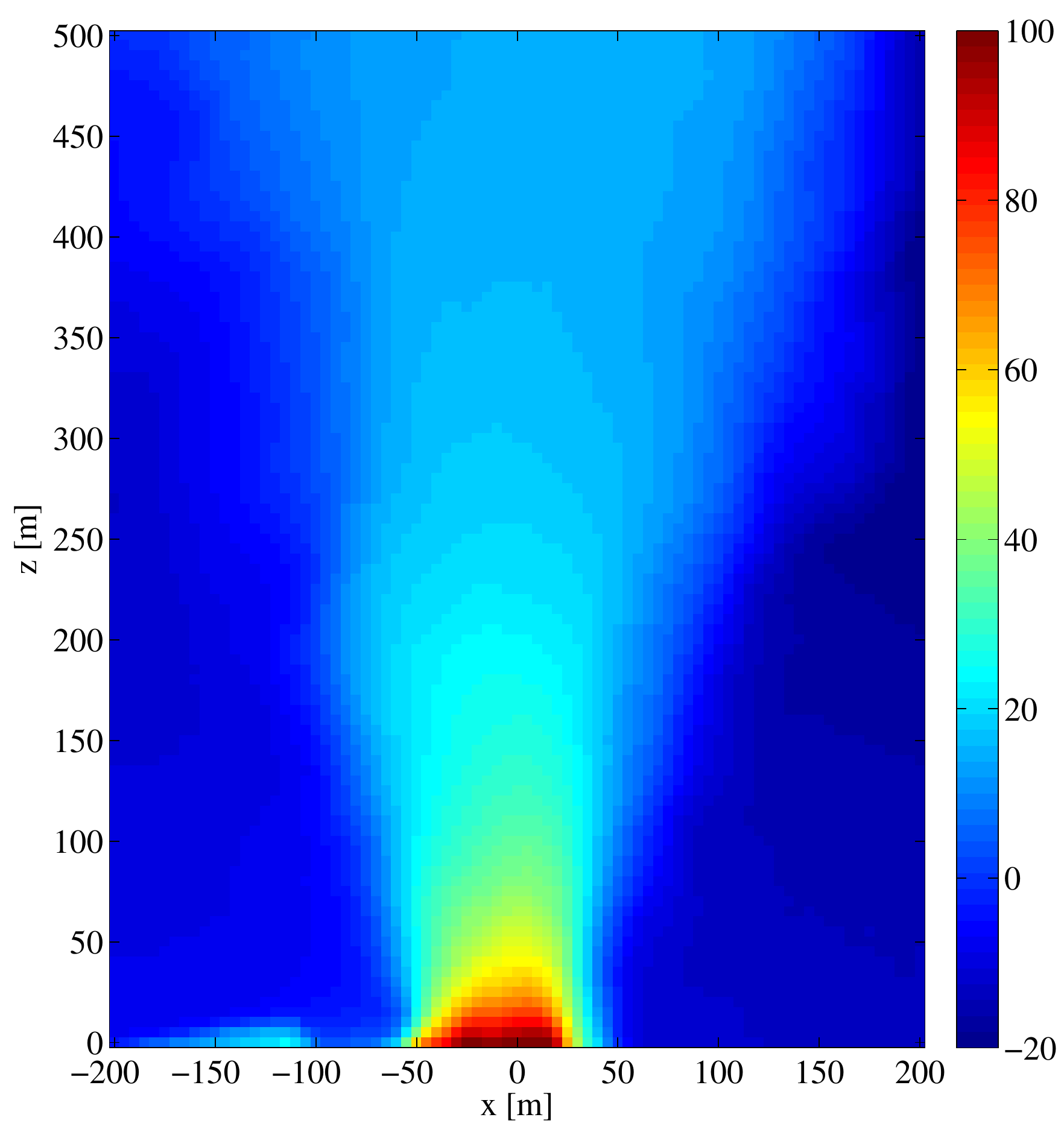}}
\caption{a) Image of the atmosphere above the volcano before the eruption. b) Averaged image of the
volcanic eruption. In both images, horizontal and vertical axes represents the position in meters
inside the image, the temperature is represented by the color scale in Celsius degrees.}
\label{fig:experiment}
\end{figure}

To analyze the TIR images, we have extracted from the full TIR dataset \citep{valade2014} a subset
in which the plume can be considered as stationary and fully developed, which starts $t_{\rm
init}=45$ s after the beginning of the eruption and ends at $t_{\rm final}=255$ s. The time-averaged
image is thus calculated (Fig.~\ref{fig:meanlmg}) and the temperature values are sampled along the
axis, at the points represented by the red dots in panel b of Fig.~\ref{fig:meanlmg}.
We have then dilated the image to partially correct the error due to the camera inclination.
Finally, we have identified a region of about 500 m in width (bounded by the horizontal dashed
lines in Fig.~\ref{fig:meanlmg}) where the flow is stationary. The image is then rotated in order to
have the plume axis along the $z$ direction. The resulting image $T_E(x_i,z_j)$ is shown in
Fig~\ref{fig:experiment}a).
Executing the same operation to the image acquired before the eruption we obtain a matrix within which we have $I_0(x_i,z_j)$ (Fig~\ref{fig:experiment}b).

\subsection{Two-dimensional inversion procedure}
We here present two possible procedures to best-fit the experimental image $I_E(x,z)$ with the
synthetic image $I_L(x,z)$ produced by the coupled fluid-electromagnetic model.
The first method is based on the two-dimensional fit to the thermal image of Figure \ref{fig:experiment}b.
Because thermal images are already converted into temperature images, we convert the synthetic intensity image $I_L$ into a thermal image $T_L(x_i,z_j)$ by using Eq.~\eqref{eq:blambda} with $\epsilon = 1$ (as in camera settings)

\begin{figure}
\centering
\subfloat[][]{\includegraphics[width=0.45\columnwidth]{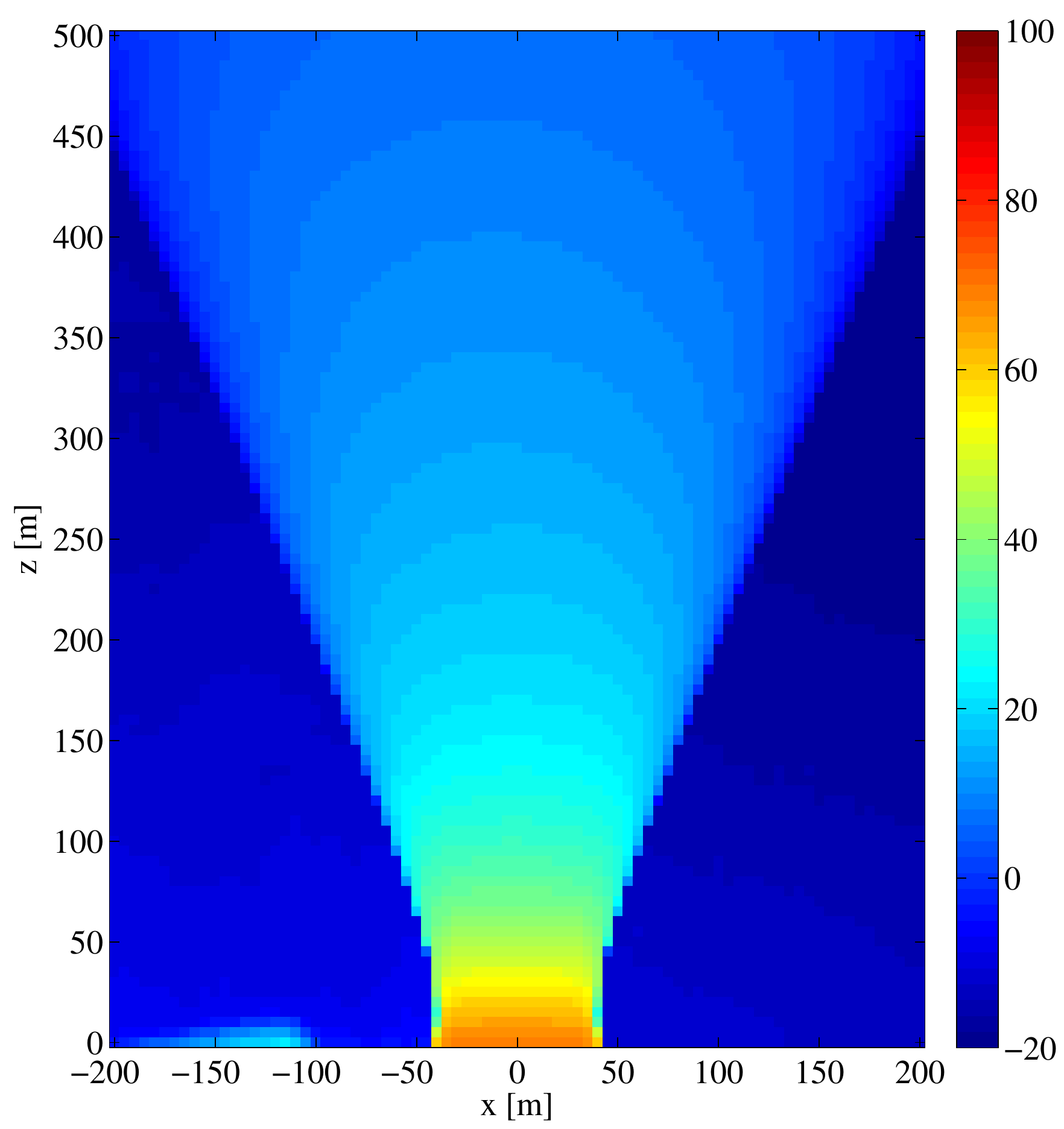}}\quad
\subfloat[][]{\includegraphics[width=0.45\columnwidth]{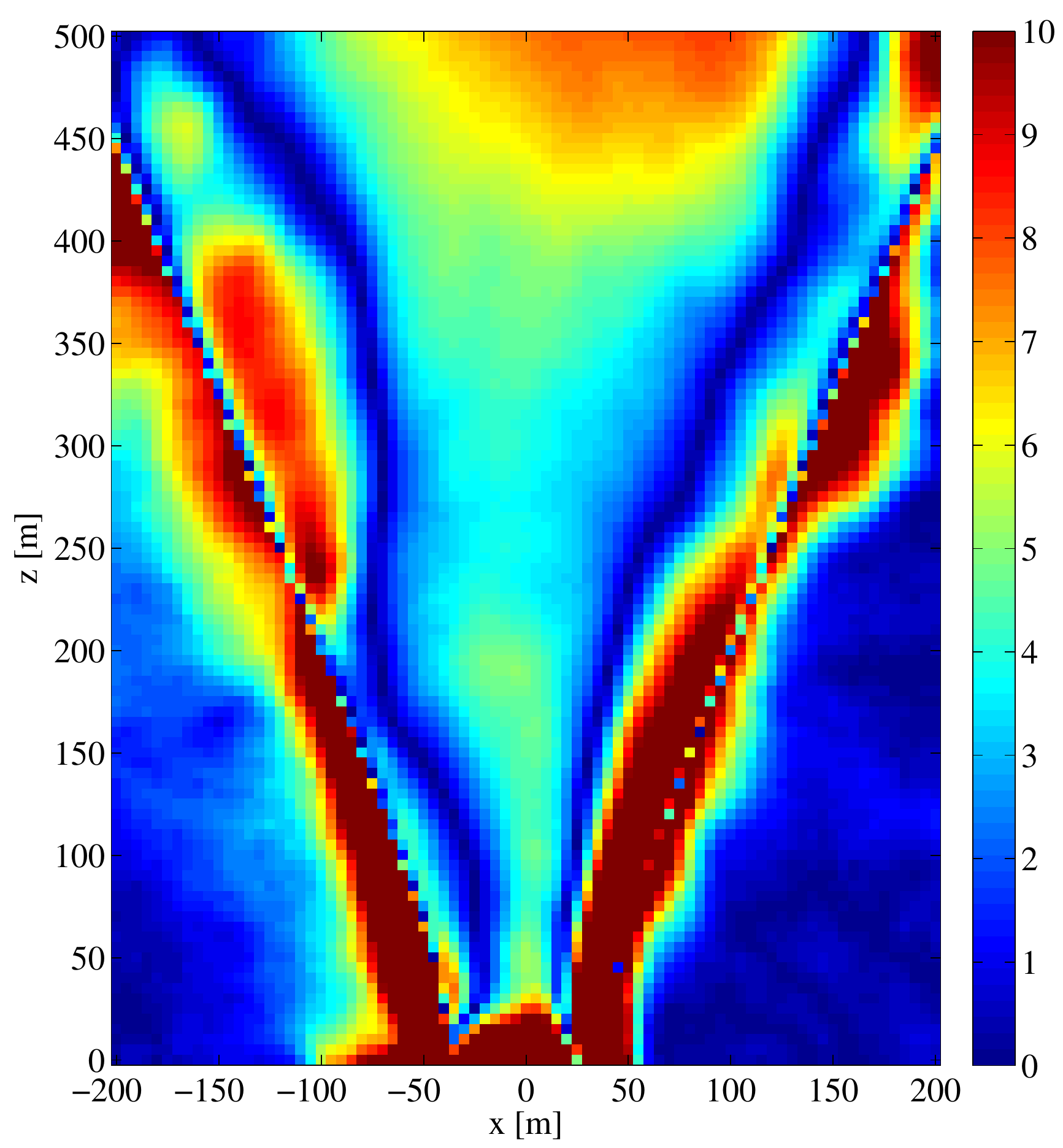}}
\caption{a) Synthetic image of the plume obtained by the two-dimensional fit of Fig.~\ref{fig:experiment}b; b) Unsigned difference between the 2D synthetic and the experimental images allowing error quantification and localization. In both images, horizontal and vertical axes represents the position (in meters) inside the image. The color scale represents the temperature in degrees Celsius.}
\label{fig:fit2D}
\end{figure}

Inversion is achieved by seeking the minimum of a cost function which measures the difference between the synthetic and the experimental images. To this end, we have chosen the following residual function:
\begin{equation}\label{eq:sigma2d}
\sigma^2(\mathbf{p}) =
\frac{1}{N*M-N_p}\,\displaystyle{\sum_{i=1}^N\sum_{j=1}^M}\left(T_E(x_i,z_j)-T_L(x_i,z_j;
\mathbf{p})\right)^2
\end{equation}
where $\mathbf{p} = (v_q,v_m,L,\phi,\chi,q_m,A_m)$ is the $N_p$-dimensional vector of parameters defining $\sigma^2$ (in this case, $N_p = 7$).
The function $\sigma^2$ must be minimized to obtain the vector of optimal input parameters $\mathbf{p} = \mathbf{p}^*$ for the plume that best fits the thermal observation.
In this application, minimization is performed by deploying a genetic algorithm (implemented in
MatLab through the function {\tt ga}), but any minimization procedure can be used. In our case,
minimization both inversions have required about 50000 trials which took about 10 s on a
laptop. The best fitting plume and the difference (in degrees Celsius) between the synthetic and the
observed plume are displayed in Fig.~\ref{fig:fit2D}. The results of the minimization procedure are
also reported in Tab.~\ref{tab:fit2D}, together with the ranges of variability specified in the
search procedure. 

\begin{figure}
\centering
\includegraphics[width=0.9\columnwidth]{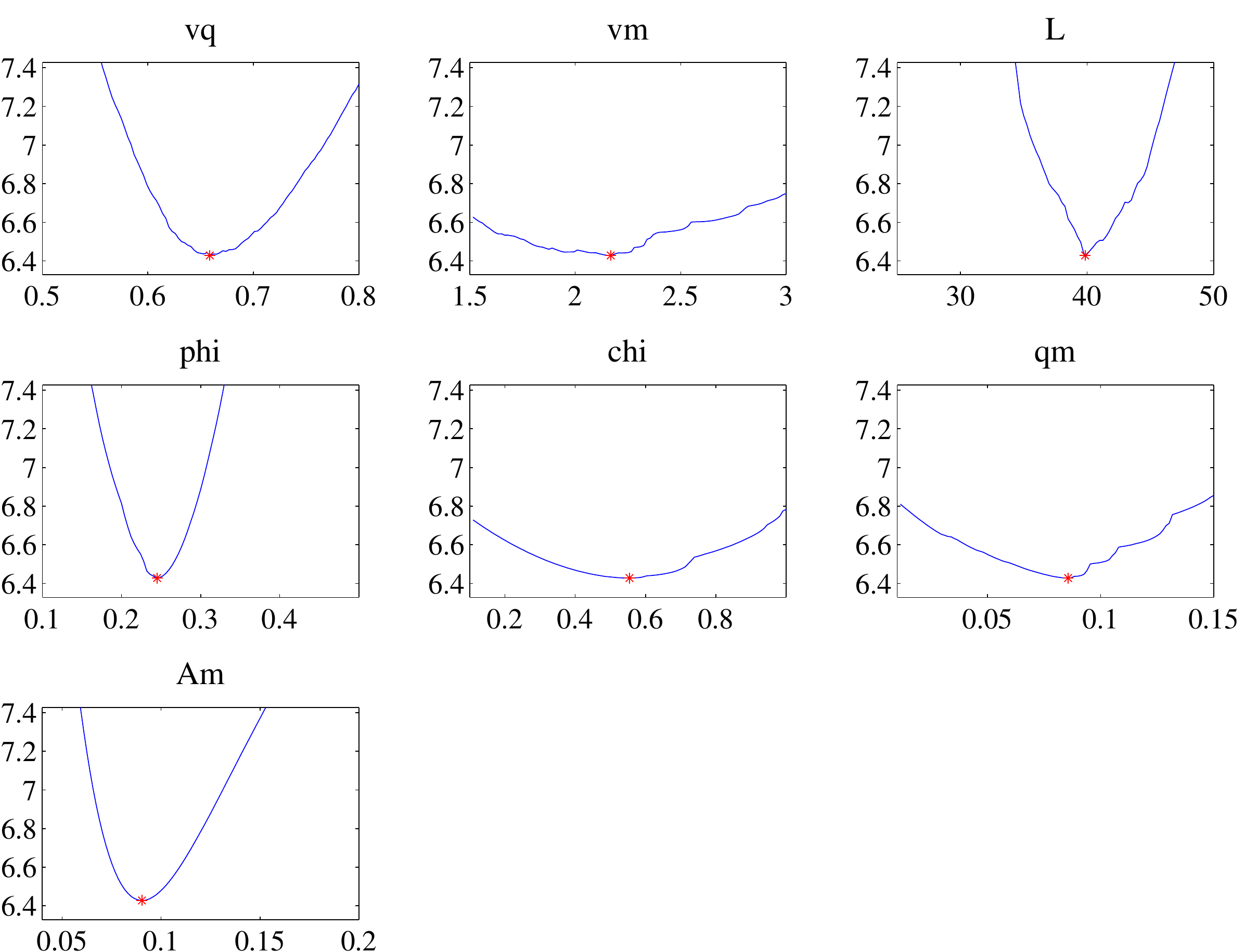}
\caption{Two-dimensional inversion procedure. Variation of  $\sigma(\mathbf{p})$ of Eq.~\eqref{eq:sigma2d} around $p^*$ as a function of each single parameter, the others being kept fixed ($\mathbf{p} = (v_q,v_m,L,\phi,\chi,q_m,A_m)$).
The red asterisk marks the value of each component of $\mathbf{p}^*$; here we obtained $\sigma(\mathbf{p}^*) = 6.428 \celsius$.}
\label{fig:err2D}
\end{figure}

%
In Fig.~\ref{fig:err2D} the projection of $\sigma(\mathbf{p}^*)$ along each parameter axis, in the
neighbours of the minimum, is shown. These plots allow to evaluate the sensitivity of the result on
the input parameters and the error associated to the solution. For this test case we obtained
$\sigma = 6.428\celsius$.

\begin{table}
\centering
\begin{tabular}{lllc}
\toprule
Parameter & Units & Range & Value \\
\midrule
$v_q$  & -- & 0.5--0.8 & $0.659\pm 0.004$\\
$v_m$  & -- & 1.5--3.0 & $2.17\pm 0.04$\\
$L$    & m  & 25--50 & $39.8\pm 0.2$\\
$\phi$ & -- & 0.1--0.5 & $0.245\pm 0.002$\\
$\chi$ & -- & 0.1--1.0 & $0.55\pm 0.02$\\
$q_m$  & -- & 0.01--0.15 & $0.086\pm 0.003$\\
$A_m$  & $\textup{m}^2/\textup{kg}$ & 0.04--0.2 & $0.0903\pm 0.0007$\\
\bottomrule
\end{tabular}
\caption{Result of the two-dimensional minimization procedure. Best fit values of the plume parameters. Here we obtained $\sigma(\mathbf{p}^*) = 6.428\celsius$.\label{tab:fit2D}}
\end{table}

\subsection{Axial inversion}
The second method is based on a one-dimensional fit of the thermal image along the plume axis. The plume axis is defined by a sequence of sampling points in the thermal image (Fig.~\ref{fig:meanlmg}a). By means of image rotation and dilation, the value of temperature along a selected region of the plume axis can be expressed as a function of the distance from the vent $T_E=T_E(z_j)$ (Fig.~\ref{fig:meanlmg}b) . 

Using only the axial points has the advantage that the background intensity is no longer important (because the plume is generally opaque along the axis) and we do not have to deal with problem of the plume edge (see the discussion below). 
However, the entrainment coefficient cannot be extracted using this procedure, so that we need a complementary analysis to evaluate its value.
To do this, we can preliminary estimate from the 2D images the plume opening angle ($\frac{{\rm d}b}{{\rm d}z}$) by defining a threshold in the temperature image \citep{valade2014}. 
The entrainment coefficient is thus derived from the plume theory \citep{ishimine2006, morton1956} as:
\[
k=\frac{5}{6} \frac{{\rm d}b}{{\rm d}z}\,.
\]
Using this method for this eruption, we obtain $k = 0.24$. 
Alternatively, we can utilize the entrainment coefficient obtained from the two-dimensional fit (Tab.~\ref{tab:fit2D}) $k = v_q/2 \simeq 0.329$.

\begin{figure}
\centering
\includegraphics[width=0.9\columnwidth]{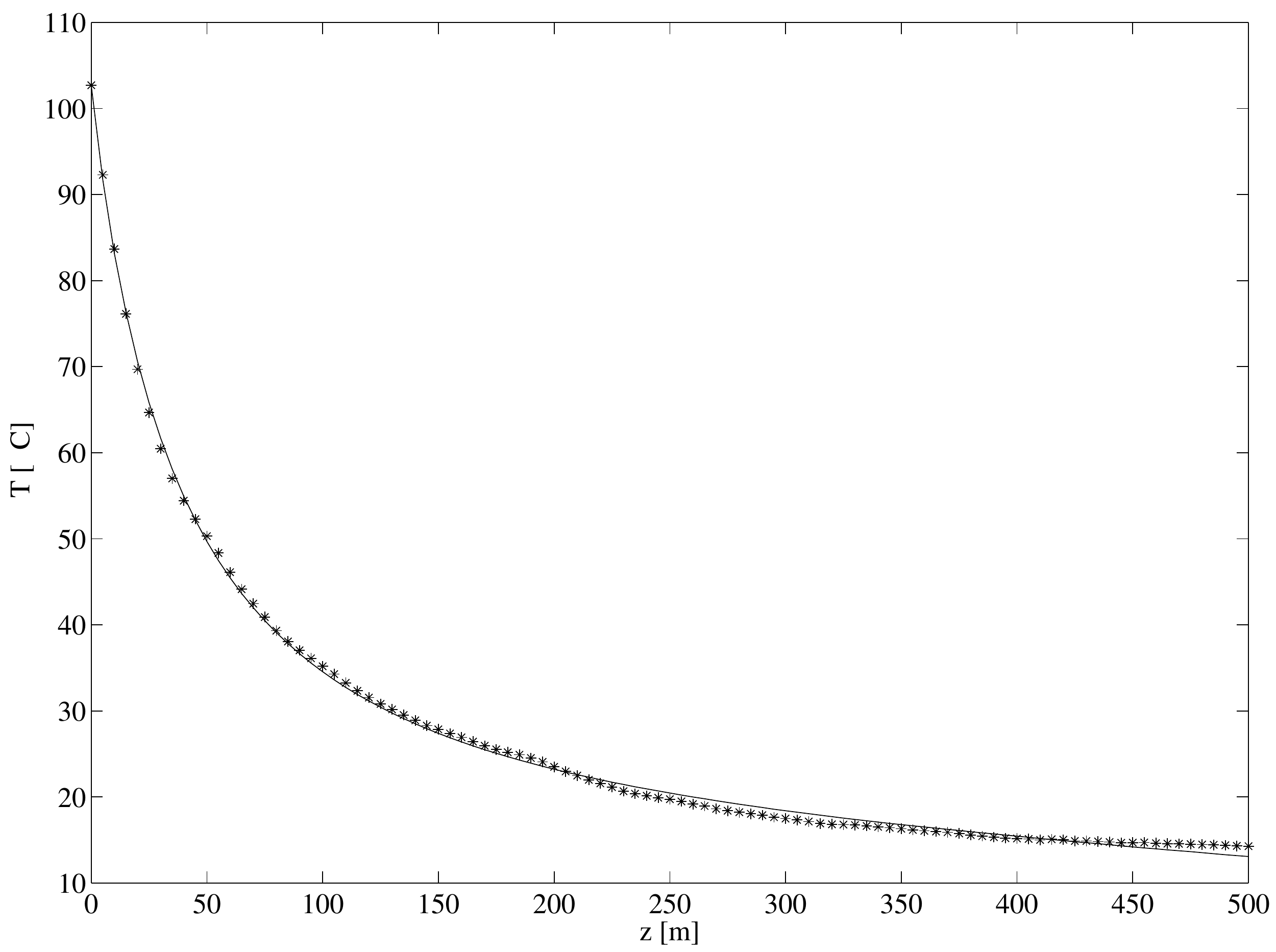}
\caption{Result of the one-dimensional fit (solid line) of the experimental thermal image along the axis (stars).}
\label{fig:1dfit}
\end{figure}

Subsequently, as for in the two-dimensional case, the synthetic temperature profile $T_L(z)$ is derived from $I_{\lambda}$ by means of Eq.\eqref{eq:blambda}. Since $k$ is independently estimated, the residual function become
\[
\sigma^2(\mathbf{p}) = \frac{\sum_{j=1}^N(T_E(z_j)-T_L(z_j;\mathbf{p}))^2}{N-N_p}
\]
where now $\mathbf{p} = (v_m,L,\phi,\chi,q_m,A_m)$ and $N_p = 6$. The result of the minimization of this new cost function is displayed in Fig.~\ref{fig:1dfit} where the fitting function (solid line) is compared with the experimental thermal data (stars). The results of the minimization procedure are also reported in Tab.~\ref{tab:fit1D}, together with the ranges of variability specified in the search procedure. 

\begin{figure}
\centering
\subfloat[][]{\includegraphics[width=0.45\columnwidth]{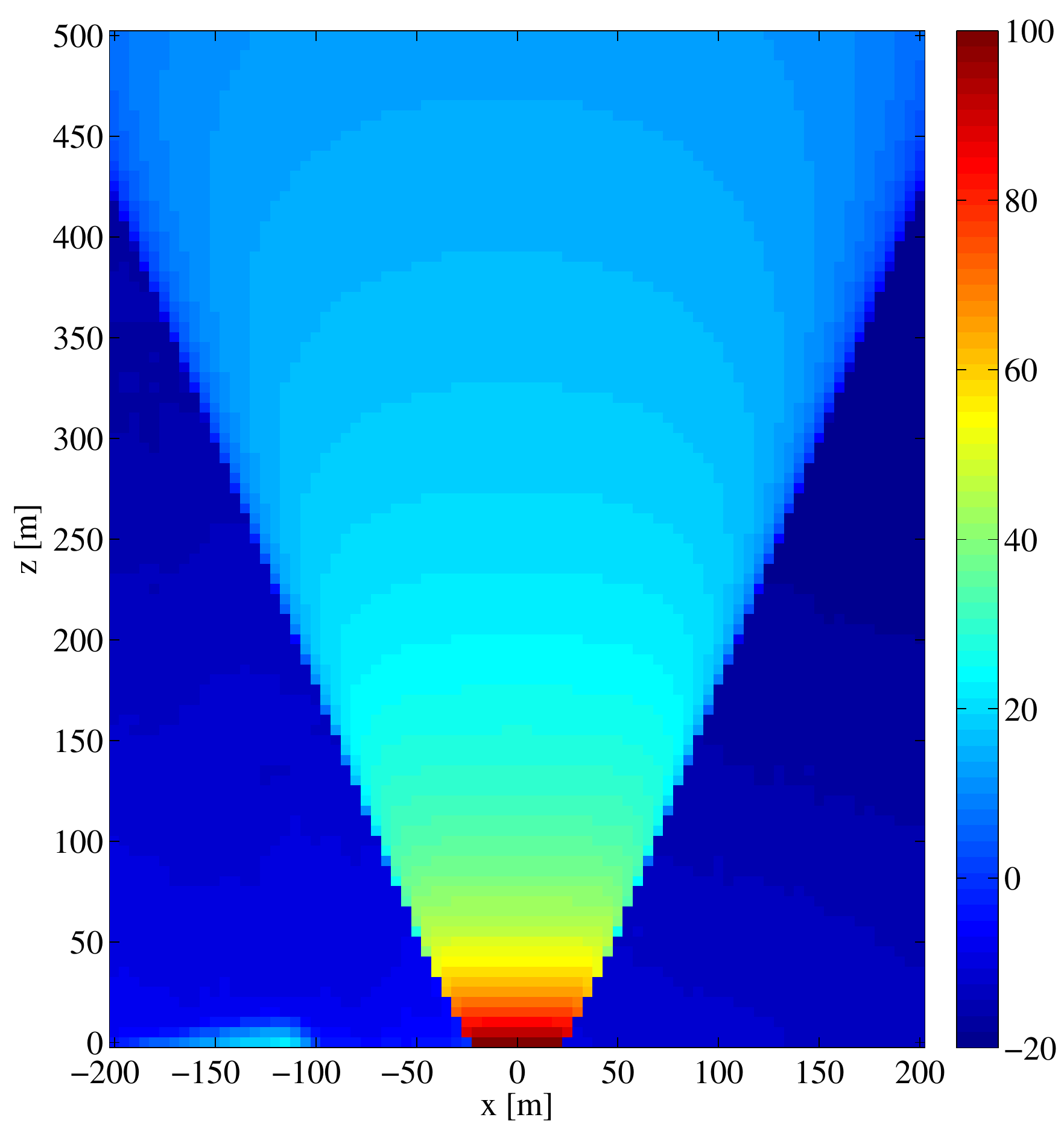}}\quad
\subfloat[][]{\includegraphics[width=0.45\columnwidth]{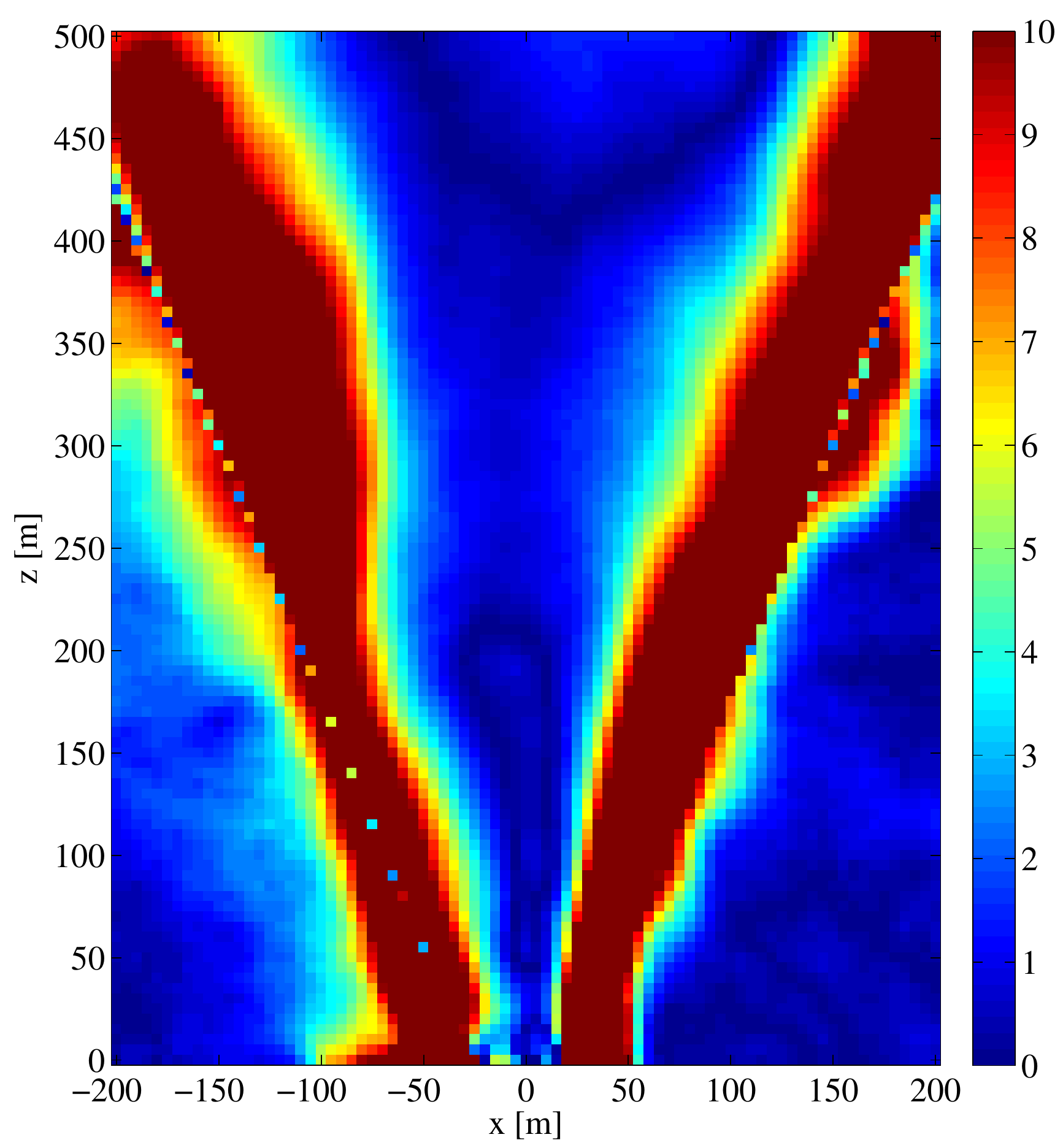}}
\caption{a) Synthetic image of the plume obtained by reconstructing the two-dimensional image from the one-dimensional fit of the axial values in Fig.~\ref{fig:experiment}b; b) Unsigned difference between the synthetic and the experimental images allowing error quantification and localization. In both images, horizontal and vertical axes represents the position (in meters) inside the image. The color scale represents the temperature in degrees Celsius.}
\label{fig:fit2D_1D}
\end{figure}

%
The corresponding 2D image, constructed by applying the top-hat profile to the one-dimensional plume model, and the difference between the optimal synthetic and the experimental images are displayed in Fig.~\ref{fig:fit2D_1D}.

\begin{figure}
\centering
\includegraphics[width=0.9\columnwidth]{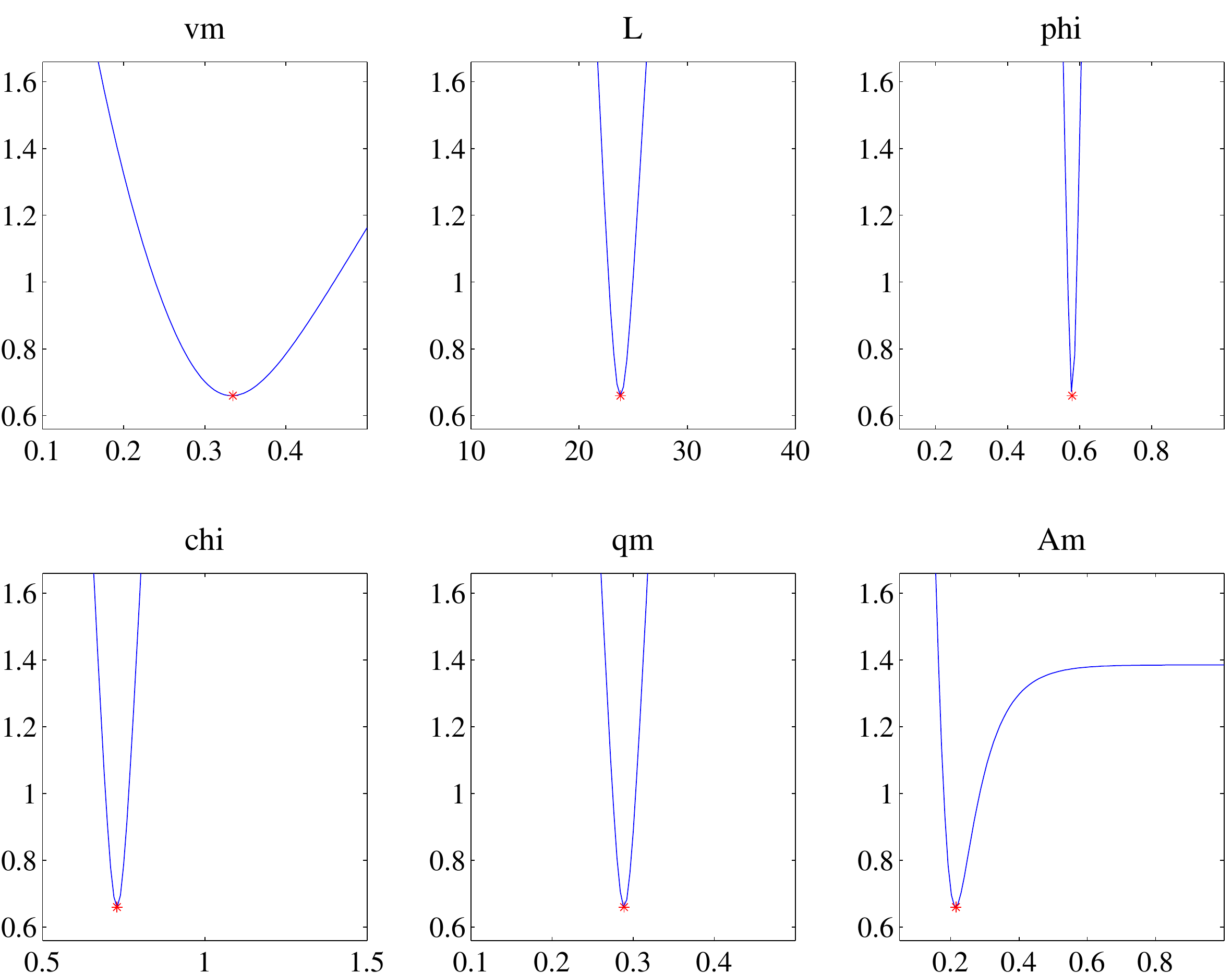}
\caption{One-dimensional inversion procedure. Variation of $\sigma(\mathbf{p})$ of Eq.~\eqref{eq:sigma1d} around $\mathbf{p}^*$ as a function of each single parameter, the others being kept fixed ($\mathbf{p} = (v_q,v_m,L,\phi,\chi,q_m,A_m)$).
The red asterisk marks the value of each component of $\mathbf{p}^*$; here we obtained $\sigma(\mathbf{p}^*) = 0.6596\celsius$.}
\label{fig:err1D}
\end{figure}

In Fig.~\ref{fig:err1D} we show the projection of $\sigma(\mathbf{p}^*)$ along each parameter axis, in the neighbours of the minimum. The error in this case is significantly reduced and all the parameters seem to be better constrained, as also indicated by the much lower value of $\sigma$, which, for this test case, is $\sigma = 0.6596\celsius$.
\begin{table}
\centering
\begin{tabular}{lllc}
\toprule
Parameter & Units & Range & Value \\
\midrule
$v_q$  & -- & -- & $0.659\pm 0.004$ \\
$v_m$  & -- & 0.1--0.5 & $0.34\pm 0.02$\\
$L$    & m  & 10--40 & $23.8\pm 0.3$\\
$\phi$ & -- & 0.1--1.0 & $0.579\pm 0.003$\\
$\chi$ & -- & 0.5--1.5 & $0.73\pm 0.01$\\
$q_m$  & -- & 0.1--0.5 & $0.29\pm 0.04$\\
$A_m$  & $\textup{m}^2/\textup{kg}$ & 0.1--1.0 & $0.215\pm 0.009$ \\
\bottomrule
\end{tabular}
\caption{Result of the one-dimensional minimization procedure. Best fit values of the plume
parameters. Here we obtained $\sigma(\mathbf{p}^*) = 0.6596\celsius$.\label{tab:fit1D}}
\end{table}

Finally, the plume input parameters (as obtained by the transformations of \ref{parinv}) are reported in Tab.~\ref{tab:input_parameters}.
As a result of the inversion procedure, we are thus able to constrain the eruption mass flow rate (in the stationary regime) as $\dot{m} = \pi Q = \pi b_0^2 \beta_0 U_0$.
The total erupted mass $m$ can be obtained by assuming a linear increase of the mass eruption rate between the eruption start and the time $t_{\rm init}$ at which the eruption is stationary. Analogously, a linear decrease of the mass eruption rate is assumed between the time $t_{\rm final}$ and the end of the eruption. Accordingly, $m=\left[\Delta t + (t_{\rm final}-t_{\rm init})\right] \times \pi Q / 2$. In order to evaluate its error, in Tab.~\ref{tab:input_parameters} we used an error on $t$ equal to 10 s.

\subsection{Parameters error estimate}
In order to give a quantitative estimation of the standard error of all the parameters, once we have
found $\mathbf{p}^*$ such that $\sigma(\mathbf{p}^*)$ is minimum, we assume that the model can be
linearized around that $\mathbf{p}^*$. In other words, naming $T_i(\mathbf{p})$ the vector of all
the measurements, we suppose that its derivative do not depends on the parameters: $\partial_{p_k}
T_i = Y_{i,k}$. In such a way, as usually done in the regression analysis \citep{bates1988}, it is
possible to formally evaluate all the fit unknowns. In particular, by using Eq.~\eqref{eq:sigma2d}
with $T_L(x_i,y_j) = T_l$ (for $i=1,\dots,N$; $j=1,\dots,M$; and $l=1,\dots,N*M$) we find
$(Y^T\,Y)_{i,j} = (N*M - N_p)\,\sigma(\mathbf{p}^*)\,\partial_{i,j}\sigma$ (for $i,j=1,\dots,N_p$).
It is worth noting that we calculate $\partial_{i,j}\sigma$ by fitting the surface $\sigma =
\sigma(\mathbf{p})$ with a second order polynomial. 

By using the classical formula for the standard error of the parameters, $\textup{se}(\mathbf{p}_i)
= \sigma(\mathbf{p}^*)\,\sqrt{(Y^T\,Y)^{-1}_{i,i}}$ and by means of error propagation, we are able
to find the confidence interval of all the parameters involved in both the axial and two-dimensional
fit (reported in Tables~\ref{tab:fit2D},~\ref{tab:fit1D}~and~\ref{tab:input_parameters}).
%
%
\begin{table}
\centering
\begin{tabular}{llcc}
\toprule
Parameter & Units & Axial fit & 2D fit \\
\midrule
$\gamma$  & --  & $0.862\pm 0.1$ & $0.543\pm 0.04$\\
$b_0$ & m & $23\pm 1$ & $41.5\pm 0.3$ \\
$Q_0$ & $10^3\,\textup{kg}/\textup{s}$  & $4.1\pm 0.5$ & $6.9\pm 0.3$\\
$M_0$ & $10^4\,\textup{kg}\,\textup{m}/\textup{s}^2$  & $3.1\pm 0.7$ & $3.1\pm 0.2$\\
$U_0$ & $\textup{m}/\textup{s}$ & $7.5\pm 0.9$ & $4.5\pm 0.2$\\
$T_0$ & $\celsius$  & $103\pm 3$ & $69.4\pm 0.3$\\
\midrule
$n_\textup{air}$ & wt.\%  & $40\pm 6$ & $85\pm 6$\\
$n_w$ & wt.\%  & $20\pm 3$ & $4.2\pm 0.3$\\
$n_s$ & wt.\%  & $41\pm 6$ & $11.1\pm 0.7$\\
$d_s$ & mm  & $3\pm 1$ & $2.1\pm 0.6$\\
\midrule
$\dot{m}_w$ & $10^3\,\textup{kg}/\textup{s}$  & $2.5\pm 0.7$ & $0.9\pm 0.1$\\
$\dot{m}_s$ & $10^3\,\textup{kg}/\textup{s}$  & $5\pm 1$ & $2.4\pm 0.3$\\
$m_w$ & $10^5\,\textup{kg}$  & $6\pm 2$ & $2.3\pm 0.4$\\
$m_s$ & $10^5\,\textup{kg}$  & $13\pm 3$ & $6\pm 1$\\
\bottomrule
\end{tabular}
\caption{Result of the 1D and 2D minimization procedure: physical input parameters for the coupled model. $m_e$ and $m_s$ are the total erupted gas and solid masses.\label{tab:input_parameters}}
\end{table}
%
\section{Discussion}
%
Comparison of the synthetic images obtained from 2D image fitting (Fig.~\ref{fig:fit2D}) and 1D fitting (Fig.~\ref{fig:fit2D_1D}) with the experimental averaged image (Fig.~\ref{fig:experiment}), shows that both inversion procedures have their maximum error along the plume boundaries.
This is due to the a-priori assumption of a top-hat self-similar profile. This assumption is accurate enough to describe the one-dimensional plume dynamics but is not accurate near the plume margins, where a Gaussian distribution better describes the actual profile.
This error is augmented in the coupled model by the fact that 1) the IR absorption depends on the density distribution, so that the top-hat model overestimates the optical thickness near the plume margins and 2) the top-hat model predicts a higher temperature on the plume margins, with respect to the Gaussian distribution. As a consequence, both effects produce a synthetic image displaying higher temperature at the margins.
To minimize this error, the 2D inversion procedure (which considers all pixels) underestimates the axial temperature (and the density) to try to balance the overestimates on the margins. In particular, at $z=0$ the error is larger because of the larger temperature contrast at this location.
This argument justifies the lower values of temperature and mass flow rate reported in Tab.~\ref{tab:input_parameters} for the 2D fit with respect to 1D.

The problem associated to the top-hat assumption is reduced when we fit only the axial values, because the integral of the absorption coefficient (i.e. the optical thickness) takes into account the whole density and temperature distribution across the plume. Therefore, the error is significantly lower in a wide region around the axis, whereas larger errors are concentrated near the margins. 
The better accuracy of the 1D fitting procedure is confirmed by the observation that the error $\sigma$ is comparable to the instrumental accuracy, which is about $0.5 \celsius$. In the 2D case, the value of $\sigma$ corroborates the conclusion that the model is not fully suited to represent the 2D shape of the plume image.

The top-hat assumption is thus more satisfactory when axial inversion is performed; a more accurate description of the plume profile is required to invert the fully 2D image. While assuming a Gaussian profile would be conceptually equivalent to the adopted top-hat hypothesis, the inversion would be computationally  more intense, because the coupled model cannot be written analytically.
The above observations also allow us to assert that the electromagnetic model is accurate enough to represent the IR emission/absorption balance throughout the plume and the error is mainly associated to application of the oversimplified fluid-dynamic model.

Despite these differences, the results of the two procedures are coherent and indicate a mass eruption rate of $7.75 \times 10^3$ in 1D and $3.31 \times 10^3$ in 2D. The observed ash plume has a gas content at $z=0$ of 59 wt.\% in 1D and 89 wt.\% in 2D. It is worth recalling here that $z=0$ does not correspond to the actual vent level but instead to the minimum quota of the analyzed thermal image. At $z=0$ the amount of entrained air is already significant (40 wt.~\% in 1D and 85 wt.\% in 2D), and it is probably associated to the specific behavior of the ash eruptions at Santiaguito which occur through a circular annulus surrounding the conduit plug \citep{bluth2004,sahetapy2009}. This can be responsible for the efficient entrainment of atmospheric air, a modification of the gas thrust region and the rapid lowering of the mixture temperature (about $100 \celsius$ in 1D and $70 \celsius$ in 2D).
The mass fraction of erupted gas (here assumed to be only water vapor), with respect to the total erupted mass, is in any case fairly high (32 wt.~\% in 1D and 27 wt.~\% in 2D).
The low ash-to-air mass ratio recovered from the inversion model are likely to be the result of emission through a porous system of cracks forming a circular vent structure \citep{bluth2004,sahetapy2009b}, so that air becomes entrapped within the ``empty'' center of the emission, thereby increasing the amount of air ingestion over cases where it enters only across the plume outer surface. The resulting ash-to-air mass ratio is low, so that the plume is dominated by heated air, with a very minor dense ash component, enhancing the buoyancy and explaining why an explosion of such low violence (mean at-vent velocities being just ~25 m s-1) can ascend to heights of between 2 and 4 km above the vent.

Finally, the estimated Sauter diameter is also comparable in the two procedures. To compare the reported values with field observations, by assuming a lognormal particle distribution with $\sigma_{\rm gsd}=1.225$ (based on the range $275-950\,\mu$m found by~\citet{wilson1980} from insitu plume sampling on filters from aircraft) and by using Eq.~\eqref{eq:d_s} we find the mean particle diameter $\bar{d} = 768\,\mu$m (in 1D) and $\bar{d} = 507\,\mu$m (in 2D).
\subsection{Plume color and visibility}
It is worth noting that in Fig.~\ref{fig:waterAbC} there are some wavelength in the visible spectral window ($\lambda < 780\,\mu$m) where the absorption coefficient of atmospheric water vapor at standard density reaches 1 m$^{-1}$, comparable with $A_w$ in the IR wavelength window considered in this paper. Moreover, the specific absorption of the ash particles is also of the same order of magnitude, $A_s\simeq 1\,\textup{m}^2/\textup{kg}$, because the assumption $d \gg \lambda$ we used to evaluate it is even more satisfied in the visible waveband.
Therefore, we can roughly say that 1) a high-temperature water-ash plume can be ``viewed'' by a thermal camera in the 8-14 $\mu$m waveband if we can see it with our eyes 2) a high-temperature water-ash plume that is opaque to our eyes is also opaque to the thermal camera 3) the plume optical thickness will be dominated by the water if $n_w \gg n_s$ or by the particles if $n_w \ll n_s$. As suggested by intuition, in the former case we will see a ``white'' plume, in the latter a black plume. Obliviously, in intermediate conditions we will see a lighter or darker gray. Now, looking at eruptions occurred at Santiaguito, the plume often appear quite light. This observation supports the argument that the erupted mixture has a quite high concentration of water.

\section{Conclusion}
\label{conclusion}
%
%
We have developed an inversion procedure to derive the main plume dynamic parameters from a sequence of thermal infrared (TIR) images.
The procedure is based on the minimization of the difference between the time-averaged experimental image and a synthetic image built upon a fluid-dynamic plume model coupled to an electromagnetic emission/absorption model.
The method is general and, in principle, can be applied to the spatial distribution of particle concentration and temperature obtained by any fluid-dynamic model, either integral or multidimensional, stationary or time-dependent, single or multiphase.

The fluid-dynamic model adopted in this work assumes self-similarity of the plume horizontal profile and kinetic and thermal equilibrium between the gas and the particulate phases. It is solved analytically in the approximation of negligible influence of atmospheric stratification (which holds below the maximum plume height considered here) and small density contrast (valid far enough above the vent) to obtain the flow parameters as a function of the height $z$ above the vent.
The electromagnetic model computes the radiated intensity from a given spatial distribution of particle concentration and temperature, assuming that particles are coarser than the radiation wavelength (about 10 $\mu$m) and neglecting scattering effects. 
In the approximation of a homogeneous plume (top-hat profile) and negligible atmospheric absorption, the coupling of the two models can be achieved analytically and allows construction of a synthetic TIR image of a volcanic plume starting from a set of vent conditions, namely the vent radius $b_0$, velocity $U_0$, temperature $T_0$, gas mass ratio $n_0$, entrainment coefficient $k$ and the equivalent Sauter diameter $d_s$ of the particle size distribution. The latter is equivalent to the average particle diameter of a monodisperse distribution having the same surface area (across a section) of the polydisperse cloud.
The minimization method, in the case analyzed here, can be applied to the single image obtained by time-averaging the sequence of TIR observations and the inversion can be performed in a few seconds on a laptop by exploiting any standard non-linear fitting software.

%
%
A test application to an ash eruption at Santiaguito demonstrates that the inversion based on the one-dimensional axial fit is preferable, because the error entailed in the inversion procedure is lower.
This is due to the top-hat assumption which overestimates density and temperature at the plume margins.
The key eruption parameters of the observed eruption are obtained, namely the mass flow rate of the gas and particulate phases, the mixture temperature and the mean Sauter diameter of the grain size distribution.\\

%
The method developed here to recover ash plume properties is fast and robust. This suggests its
potential applications for real-time estimation of ash mass flux and particle size distribution,
which is crucial for model-based projections and simulations. By streaming infrared data to a
webtool running, in real-time, the model could provide the input parameters required for ash
dispersion models run by VAACs.

The algorithm could also be easily applied to more complex geometric configuration (e.g., to a bent plume in a wind field -- \citet{woodhouse2013}) and atmospheric conditions (e.g., in presence of a significant amount of water vapour), or to more realistic plume models (e.g., assuming a Gaussian plume profile). In such cases, the coupled model should be solved numerically. 
It is worth noting that the calibration of the background atmospheric infrared intensity and the information on the atmospheric absorption can be critical in the applications and we recommend experimentalists to consider their effects during the acquisition campaigns.
Also, the intensity image would be preferable with respect to the temperature image, which is derived from automatic onboard processing by commerical thermal cameras. Finally, it is worth noting that a rigorous validation of the direct model (i.e., the generation of the synthetic image) must still be achieved. Unfortunately, we could not find detailed experimental measurements of the TIR radiation from a turbulent gas-particle plume under controlled injection conditions. This would be extremely useful to calibrate the coupled forward model and to better understand plume visibility issues.

\section*{Acknowledgements}
This work presents results achieved in the PhD work of the first author (M.C.), carried out at Scuola Normale Superiore and Istituto Nazionale di Geofisica e Vulcanologia. Support by the by the European Science Foundation (ESF), in the framework of the Research Networking Programme MeMoVolc, is also gratefully acknowledged.  AH (and SV until June 2013) was supported by le R\'egion Auvergne and FEDER.
\appendix
\section{Thermodynamics of the dusty gas}
\label{thermo}
Closure of Eq. \ref{eq:Woods} requires the specification of the thermodynamics of all components of
the eruptive mixture.
We indicate with $C$ the specific heats at constant pressure, that we assume constant (see Tab.
\ref{tab:thermo}). The dusty-gas specific heat is expressed as a function of its components as:
\begin{equation}
 \Cb = \frac{1}{\beta}(\rho_s C_s + \rho_e C_e + \rho_\alpha \Ca)
\end{equation}
which, in terms of the conserved variables, can be written as:
\begin{equation}
 \Cb = \Ca + \frac{Q_s}{Q}(C_s - \Ca) + \frac{Q_e}{Q}(C_e - \Ca)\,.
\end{equation}
We indicate with $R$ the gas constants (see Tab. \ref{tab:thermo}). The dusty-gas constant is
expressed as a function of its components as:
\begin{equation}
 R_g = R_\alpha + \frac{\rho_e}{\beta - \rho_s} (R_e - R_\alpha) = R_\alpha + \frac{Q_e}{Q - Q_s}
(R_e - R_\alpha)\,.
\end{equation}
The following quantities are also defined:
$\chi_s = C_s/\Ca$, $\chi_e = C_e/C_a$, $\psi_e = R_e/R_\alpha\,.$

Finally, following \citet{woods1988}, the density $\beta$ of the dusty gas is expressed as a
function of the conserved variables as:
\begin{equation}\label{eq:beta-1_1}
 \frac{1}{\beta} = \frac{1 - n}{\sigma} + \frac{n R_g \Tb}{p} = \frac{1}{\hat{\rho}_s} \frac{Q_s}{Q}
+
\left(1- \frac{Q_s}{Q}\right) \frac{R_g \Tb}{p}\,,
\end{equation}
where the particle (average) material density is taken as constant $\hat{\rho}_s = 1600 {\rm
kg/m}^3$.
\begin{table}
\centering
\begin{tabular}{ccc}
\toprule
Gas species & $R$ & $C_P$ \\
\midrule
Water vapor              & 462 & 1862 \\
Atmospheric air          & 287 & 998 \\
Ash                      &  -- & 1100 \\
\bottomrule
\end{tabular}
\caption{Thermo-physical parameters of the plume components.}
\label{tab:thermo}
\end{table}

\section{Variable transformation}
\label{nondim}

The dimensioned variables can be expressed in terms of the non-dimensional variables $(q,m,f)$ by
the following transformations:
\begin{eqnarray}
\label{eq:betadiq}
\beta&=&\alpha\frac{q (q + \chi q_m)}{(\phi+q)(q-q_m)}\\
\label{eq:bdiq}
b&=&L\sqrt{\frac{\alpha_0}{\alpha} \frac{q(\phi+q)(q-q_m)}{m (q + \chi q_m)}}\\
\label{eq:udiq}
U&=&U_0 \frac{m}{q}\\
\label{eq:tbetadiq}
T_{\beta}&=&T_{\alpha} \frac{\phi+q}{q+\chi q_m}
\end{eqnarray}

\section{Parameter inversions}
\label{parinv}
The solution of the plume model Eq.~\ref{eq:analytical} in non-dimensional form is function of the boundary values and model parameters, $(v_q, v_m, q_m, \phi, \chi, L)$.
The inversion procedure described in Sect.~\ref{inversion} provides the set of parameters which minimizes the difference between the synthetic and the exprimental image. We here report the transformation needed to obtain the equivalent set of eruption parameters. 

We first use Eq. \eqref{eq:vm} together with the expression for $L=Q_0/\sqrt{\alpha_0 M_0}$ to obtain:
\begin{eqnarray}
\label{eq:M0}
M_0 &=& \frac{g \phi \alpha_0}{v_m}(1-\gamma)\\
\label{eq:Q0}
Q_0 &=& \sqrt{\alpha_0 M_0} L\\
\label{eq:U0}
U_0 &=& \frac{M_0}{Q_0}
\end{eqnarray}
By means of Eqs. \eqref{eq:bdiq} and \eqref{eq:tbetadiq}, putting ($q=1$, $m=1$) and $\alpha=\alpha_0$ at $z=0$, we also obtain:
\begin{eqnarray}
\label{eq:b0}
b_0 &=& \sqrt{\frac{(1+\phi)(1-q_m)}{1 + \chi q_m}} L\\
\label{eq:T0}
T_0 &=& T_{\alpha 0} \frac{\phi+1}{1+\chi q_m}
\end{eqnarray}

Finally, the composition of the eruptive mixture can be reconstructed by noting that:
\begin{eqnarray}
q_m &=& n_s - (\psi_w - 1) n_w \\
\chi &=& \frac{(\chi_s-1)n_s-(\chi_w-1)n_w}{q_m}
\end{eqnarray}
where we have assumed that the erupted gas is composed of water vapour (subscript $w$).
The system can be solved to obtain the mass fractions of ash, volcanic gas and atmospheric gas at
$z=0$:
\begin{eqnarray}
n_w &=& \frac{(1 + \chi - \chi_s) q_m}{(\chi_s - 1)(\psi_w - 1)+ (\chi_w - 1)} \\
n_s &=& q_m \frac{\chi (\psi_w - 1) + (\chi_w - 1)}{(\chi_s - 1) (\psi_w - 1) + (\chi_w - 1)}\\
n_{\alpha} &=& 1 - n_w - n_s\,.
\end{eqnarray}
Note that, in general, the gas mass fraction in the mixture at $z=0$ is $n_0 = n_w + n_{\alpha}$.
Since the quota $z=0$ may not correspond to the vent quota (it is better defined as the quota where
the plume starts to be self-similar and stationary), $n_0$ in general does not correspond to the gas
content in the eruptive mixture but may also contain the fraction $n_{\alpha}$ of entrained air.

\begin{figure}
 \centering
 \subfloat[][]{\includegraphics[width = 0.45\columnwidth]{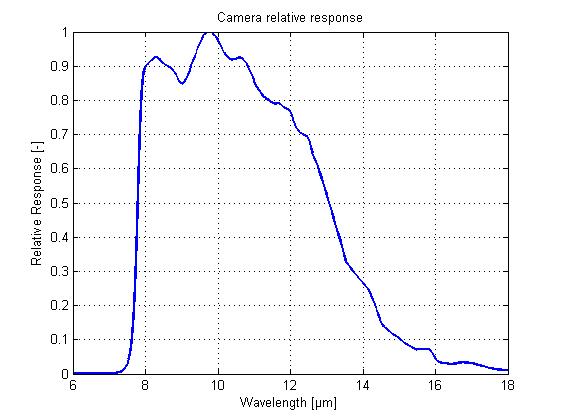}}\quad
 \subfloat[][]{\includegraphics[width=0.45\columnwidth]
 {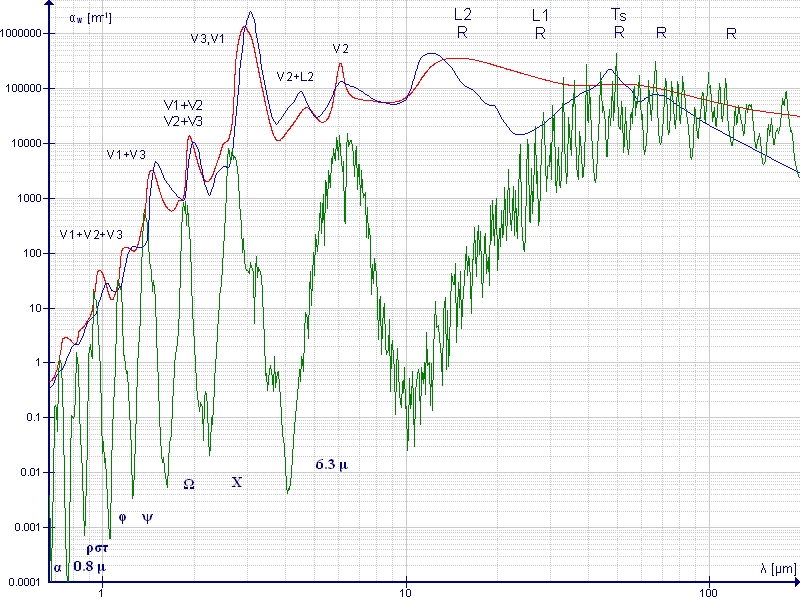}}
 \caption{a) An example of a typical spectral response of a FLIR camera in the spectral window
$7\div14\,\mu\textup{m}$. b) {\em From Wikipedia:} Absorption spectrum (attenuation coefficient vs.
wavelength) of liquid water (red), atmospheric water vapor (green) and ice (blue line) between 667
nm and 200 $\mu$m. The plot for vapor is a transformation of data {\em Synthetic spectrum for gas
mixture 'Pure H2O'} (296K, 1 atm) retrieved from Hitran on the Web Information System.}
 \label{fig:waterAbC}
\end{figure}

Finally, the equivalent Sauter diameter of the grain size distribution can be derived by the
absorption coefficient $A_m$ by assuming that the absorption by atmospheric air is negligible. In
such case, in Eq.~\eqref{eq:kmixdiq} the specific absorption coefficient can be written as:
\[
 A_m = A_s q_s + A_w q_w = A_s n_s + A_w n_w\,.
\]
Therefore, knowing the specific absorption coefficient of water vapour $A_w$ and the components mass fraction, the specific absorption coefficient for particles $A_s$ can be derived from this expression. By noting that $\displaystyle A_s = \frac{3}{2 d_s \hat{\rho}_s}$, this can be used to estimate $d_s$.
Please notice that the absorption coefficient for water should be estimated by the experimentalists as the convolution of the detector response function with the spectral absorption coefficient for the pure gas. 
For example, in the case of atmospheric water vapour, in order to evaluate $A_w$, we have to execute the convolution of the spectral response of our camera (cf. Fig.~\ref{fig:waterAbC}a for an example) and the absorption spectrum of the atmospheric water vapour (Fig.~\ref{fig:waterAbC}b). Throughout the paper we used the value $A_w = 1 \textup{m}^2/\textup{kg}$ that, as we can see in Fig.~\ref{fig:waterAbC}, is the right order of magnitude for the absorption coefficient of atmospheric water vapour.

\bibliographystyle{model2-names}
\bibliography{biblio}

\end{document}